\documentclass[10pt,preprint]{aastex}
\usepackage{graphicx}
\usepackage{amsmath}
\usepackage{mathtools}
\usepackage{ wasysym }

\def\orch{\textit{Orchestra}}
\def\orchestra{\orch}
\def\nbody{$n$-body}

\def\Msolar{{M$_\odot$}}
\def\Mp{M_p}
\def\Ms{M_s}
\def\mreduced{\mu}
\def\Mfac{{\cal M}}
\def\Rp{R_p}
\def\Rs{R_s}
\def\phip{\phi_\text{pri}}
\def\phis{\phi_\text{sec}}
\def\abin{a_\text{bin}}
\def\ebin{e_\text{bin}}
\def\nbin{\Omega_\text{bin}}
\def\nsyn{\Omega_\text{syn}}
\def\nkep{\Omega_\text{Kep}}
\def\rgc{R_g}
\def\rgcest{\tilde{R}_g}
\def\drplus{\Delta R_{+}}
\def\drminus{\Delta R_{-}}
\def\drpm{\Delta R_{\pm}}
\def\rmin{R_{\text{min}}}
\def\rmax{R_{\text{max}}}
\def\ngc{\Omega_g}
\def\kappae{\kappa_e}
\def\kappai{\kappa_i}
\def\chie{\chi_e}
\def\chii{\chi_i}
\def\eest{\tilde{e}_\text{free}}
\def\efree{e_\text{free}}
\def\eforce{e_\text{force}}
\def\inc{\imath}
\def\Cj{C_J}
\def\akep{a_\text{Kep}}
\def\ekep{e_\text{Kep}}
\def\ageo{a_\text{geo}}
\def\egeo{e_\text{geo}}
\def\robs{R_\text{obs}}
\def\rddobs{\ddot{R}_\text{obs}}
\def\phiobs{\phi_\text{obs}}
\def\phidotobs{\dot{\phi}_\text{obs}}
\def\phiddobs{\ddot{\phi}_\text{obs}}
\def\thq{\text{mc}}
\def\rtheory{R_\thq}
\def\phitheory{\phi_\thq} 
\def\rddtheory{\ddot{R}_\thq} 
\def\phiddtheory{\ddot{\phi}_\thq}
\def\vkep{{v_\text{Kep}}}
\def\vrnd{{v_\text{rnd}}}

\begin{document}

\title{On the estimation of circumbinary orbital properties}

\author{Benjamin C. Bromley}
\affil{Department of Physics \& Astronomy, University of Utah, 
\\ 115 S 1400 E, Rm 201, Salt Lake City, UT 84112}
\email{bromley@physics.utah.edu}

\author{Scott J. Kenyon}
\affil{Smithsonian Astrophysical Observatory,
\\ 60 Garden St., Cambridge, MA 02138}
\email{skenyon@cfa.harvard.edu}

\begin{abstract}
  We describe a fast, approximate method to characterize the orbits of satellites around a central binary in numerical simulations. A goal is to distinguish the free eccentricity --- random motion of a satellite relative to a dynamically cool orbit --- from oscillatory modes driven by the central binary's {time-varying} gravitational potential. We assess the performance of the method using {the} Kepler-16, Kepler-47, and Pluto-Charon systems. We then apply the method to a simulation of orbital damping in a circumbinary environment, resolving relative speeds between small bodies that are slow enough to promote mergers and growth.  These results illustrate how dynamical cooling can set the stage for the formation of Tatooine-like planets around stellar binaries and the small moons around the Pluto-Charon binary planet. 
\end{abstract}

\keywords{planets and satellites: formation
-- planets and satellites: individual: Pluto}

\section{Introduction}

Planet formation seems inevitable around nearly all stars.  Even the rapidly changing gravitational field around a binary does not prevent planets from settling there \citep[e.g.,][]{armstrong2014}. Recent observations have revealed stable planets around binaries involving main sequence stars \citep[e.g., Kepler-16;][]{doyle2011}, as well as evolved compact stars \citep[the neutron star-white dwarf binary PSR~B1620-26;][]{sigurdsson2003}. The Kepler mission \citep{borucki2011} identified a dozen other close-in circumbinary planets \citep{welsh2012, orosz2012a, orosz2012b, schwamb2013, kostov2013, kostov2014, welsh2015, kostov2016}. More recently, TESS \citep{ricker2015, huang2018} detected one more addition to the census \citep[TOI-1338]{kostov2020}. Closer to our own home, the Sun's binary planet, Pluto-Charon, has a compact system of small moons \citep{christy1978, buie2006, showalter2011, showalter2012, weaver2016} that have all found a stable home close to the binary partners.

From a theoretical perspective, the gravitational perturbations from a central binary present some hurdles for the formation of circumbinary satellites --- planets around double stars, or moons around binary planets \citep[e.g.,][]{holman1999, mori2004, quintana2006, scholl2007, pierens2007, doolin2011, kennedy2012, raf2013, pierens2013, lines2014, kennedy2015, bk2015tatooine,  kley2015}.  Analytical studies \citep{sutherland2019} and extensive simulations \citep[e.g.,][]{chavez2015, fleming2018, quarles2018} highlight the importance of disruptive orbital instabilities related to resonances. Other issues affecting the orbital architecture of circumbinary planets include the initial mass distribution of protoplanetary disks around binaries, and the ability of nascent planets to migrate within these disks \citep[e.g.,][]{schlichting2014}.

The growing census of circumbinary planets demonstrates that these theoretical obstacles are surmountable. Robust processes that drive the formation of planets around a single star are likely at play in circumstellar environments: coagulation, accretion, and mergers in a swarm of planetesimals lead to growth of planets, regulated by a balance between dynamical excitation (for example, gravitational stirring of small bodies by larger ones, which causes high-speed destructive collisions), and dynamical cooling \citep[collisional damping and dynamical friction, which facilitate low-speed collisions and mergers; see][]{saf1969, weth1980, weth1993, liss1993b, ida2004}. The success of this scenario has been borne out in simulations that track the growth of planets from a sea of planetesimals \citep[e.g.,][]{spaute1991, weiden1997b, kok1995, kl1998, chambers2004, lev2012, kb2008, bk2011a}.

Dynamical cooling within a protoplanetary disk is essential to the growth of planets. Around a single star, dynamically cold orbits are circular. Particles that lie on them in a common orbital plane follow paths that are nested and never cross. Orbital eccentricity $e$, which is easily tracked in \nbody\ simulations of planet formation, is a measure of random motions relative to these trajectories; collisions in a swarm of objects with low eccentricities yield mergers, while higher eccentricities result in erosion and fragmentation. This connection between eccentricity and collision outcomes is particularly useful in simulations that determine collision rates based on tracer particles \citep[e.g.,][]{lev2012} or statistically evolved particle ensembles \citep[e.g.,][]{kl1998}.

Around a central binary, the time-varying gravitational potential makes dynamically cold orbits harder to identify. Yet they exist: "Most-circular" orbits are the analog of circular trajectories around a single central mass \citep{lee2006, youdin2012, leung2013, bk2015tatooine}. Most-circular trajectories lie in the plane of the binary; particles on these paths flow with the binary's motion but never cross the orbits of particles on adjacent paths, even when the binary is eccentric.  Random motion relative to a most-circular path is measured by the free eccentricity $\efree$ \citep[e.g.,][]{lee2006}; dynamically cold orbits have $\efree = 0$. Because of the complicated motion driven by the binary on dynamical time scales, $\efree$ is more difficult to isolate than its Keplerian counterpart $e$ even when $\efree$ is identically zero. In simulations of circumbinary planet formation, it is a challenge to distinguish between dynamically hot orbits that lead to destruction from cool ones that lead to growth.

Here, we take up this challenge, focusing on how to characterize, track, and manipulate orbits of small bodies around a central binary in dynamical simulations. The goal is to calculate realistic trajectories for these objects, accurately resolving orbital parameters, including their free eccentricity, as they settle onto most-circular orbits. Only by tracking orbits with this level of detail can we measure relative collision speeds well enough to distinguish between merger events and catastrophically destructive impacts. However, existing methods for estimating orbital elements around binaries \citep[e.g.,][]{showalter2015, woo2018} involve either model fitting and/or the storage of up to thousands of sample points along individual trajectories. These approaches are not practical in a large $n$-body simulation. Instead, the estimators we seek must be computed quickly and involve little data storage. We are not looking to replace existing estimation methods used with hard-won observational data so that we may sacrifice some accuracy for an algorithm that is fast and lean.

We begin this work with a review of the epicyclic theory of circumbinary orbits introduced by \citet{lee2006}, which is the foundation for much of our analysis (\S\ref{sec:theory}). In \S\ref{sec:orbitchar}, we outline ways to characterize circumbinary orbits and propose simple measures of circumbinary orbital elements that can be calculated efficiently from a single state vector. Focusing on the free eccentricity, we illustrate the performance of our method using several observed circumbinary systems (Kepler-16, Kepler-47, and Pluto-Charon). Next, in \S\ref{sec:comp}, we compare our estimator to other values of $\efree$ from the literature for the small moons of Pluto-Charon. Then, in \S\ref{sec:apps}, we demonstrate how to use this estimator in a simulation of orbital damping. We conclude in \S\ref{sec:conclude}.

\section{The analytical model: linearized theory}\label{sec:theory}

In planet formation simulations, dynamical friction, viscous stirring and collisional damping modify the orbital parameters of solids. Having a prescription to measure and change these orbital elements during a numerical calculation is important. With a single central object, the motion of satellites --- planets around a star, or moons around a planet --- is well-described by Keplerian orbital elements. We can then monitor quantities like orbital distance and eccentricity easily from a single state vector. Modifying orbital elements is straightforward as well \citep{bk2006}. Such changes to a satellite's trajectory in an $n$-body calculation are warranted if, for example, the satellite experiences dynamical friction from interactions with a massive swarm of small dust particles that are not explicitly evolved as $n$-bodies \citep[e.g.,][]{bk2006, kb2009}. 
 
For a satellite orbiting a binary, it is not immediately clear how to \textit{define} orbital eccentricity, much less how to modify it. Eccentricity might be associated with radial excursions from some average orbital distance, as in the Keplerian case. However, the binary's rapidly varying gravitational potential influences the extent of radial traversals; they are a mixture of random motion and forced modes that respond to the binary's orbit. In secular perturbation theory \citep[e.g.,][]{murray1999}, the rapidly varying binary potential is replaced by a time-averaged one; the free eccentricity derived from it corresponds to precessing, elliptical motion, superimposed on forced eccentric motion driven by the eccentric orbit of the binary. A satellite's random motion relative to a most-circular orbit (which includes forced eccentric motion from secular theory; \citealt{leung2013}) is characterized by the free eccentricity. However, secular theory ignores important behavior that takes place on dynamical time scales. Thus it does not provide a prescription for parameter estimation from a state vector.

As a guide out of this thicket, we adopt the epicyclic theory of \citet{lee2006}. It provides explicit orbit solutions that allow us to distinguish motion driven by the binary's time varying gravitational potential from random motion, as quantified by the free eccentricity. The Lee-Peale theory is designed for orbits that are close to circular and are nearly coplanar with the binary. Errors in orbit solutions scale as the square of the orbital inclination. Similarly, errors appear at second order in $\efree$ and (in the extension of the theory by \citealt{leung2013}) in the binary eccentricity, $\ebin$. Other theoretical solutions are known \citep[e.g.,][]{georgakarakos2015}, although we do not consider them here.

\citet{lee2006} derived equations of motion of a satellite by first designating a guiding center, a reference point in uniform circular motion that tracks the satellite's overall path around the binary.  They define coordinates relative to the guiding center location that are used to determine the satellite's position and speed as it makes excursions about this reference point. The exact equations of motion are linearized in these coordinates. The time variation in the gravitational potential experienced by the satellite during these excursions is then expressed in terms of harmonics of orbital frequencies (see Appendix~\ref{appx:linear} for details). Between the linearization in excursion coordinates and the decomposition of the time dependence into harmonic frequencies, an orbit solution emerges that is the linear superposition of harmonic oscillator modes:
\begin{eqnarray}
\label{eq:R}
  R(t) & = & \rgc\left[1 - \efree\cos(\kappae t+\chie) - 
     \sum_{k=1}^\infty {C_k}\cos(k\nsyn t)
     \right]\!, 
\\
\label{eq:phi}
  \phi(t) & = & \ngc\left[ t + \frac{2\efree}{\kappae} 
\sin(\kappae t+\chie) + 
\sum_{k=1}^\infty \frac{D_k}{k\nsyn}\sin(k\nsyn t)
\right]
\!,
\ \ \text{and}
\\
\label{eq:z}
z(t) & = & i \rgc \cos(\kappai t + \chii), \vphantom{\frac{3^3}{3_3}}
\end{eqnarray} 
where $(R,\phi,z)$ is the satellite's position at time $t$ in cylindrical coordinates, and the parameters on the right-hand sides are listed in Table~\ref{tab:vars}. Here, we choose $t = 0$ to correspond to the time when the satellite and the secondary are aligned, with azimuthal coordinates $\phi$ and $\phis$ set to zero. This orbit solution applies only for the case of a circular binary ($\ebin = 0$), as detailed in Appendix~\ref{appx:linear} following \citet{lee2006}.  \citet{leung2013} generalize for motion around eccentric binaries. 

Broadly, this orbit solution tracks the motion of the guiding center (the leading terms on the right Equations (\ref{eq:R}) and (\ref{eq:phi})), random motion (terms with free eccentricity $\efree$ and inclination $i$), and forced oscillations that result from the binary's motion (the terms in the summations with mode amplitudes $C_k$ and $D_k$). The frequencies $\kappae$ and $\kappai$, correspond to epicyclic and vertical motion, while the synodic frequency, $\nsyn$, is the rate at which the satellite is in conjunction with the binary, setting the tempo of the regular kicks that the satellite receives as it swings by.

\begin{deluxetable}{lll}
  \tablecaption{\label{tab:vars} Parameters in the orbit solutions from linearized theory}
  \tablehead{ \colhead{parameter} & \colhead{description} &
    \colhead{definition/reference}}
    \startdata
    $\Mp$, $\Ms$, $M$ & primary, secondary \& total mass & $M = \Mp+\Ms$ \\
    $\abin$, $\ebin$ & binary semimajor axis and eccentricity & --- \\
    $\nbin$ & mean motion of the binary & $\sqrt{GM/\abin^3}$ \\
    $\rgc$ & orbital radius of a satellite's guiding center & Eq.~(\ref{eq:R}) \\
    $\ngc$ & mean motion of guiding center & Eqs.~(\ref{eq:phi}) and (\ref{eq:ngc})
    \\  $\nsyn$ & synodic frequency & $\nbin-\ngc$ \\
    $C_k$, $D_k$ & amplitudes of forced oscillations & Eqs.~(\ref{eq:R}), (\ref{eq:phi}), (\ref{eq:C}), and (\ref{eq:D}) \\
    $\efree$& free eccentricity & Eq.~(\ref{eq:R})  \\
    $\kappae$ & epicyclic motion & Eqs.~(\ref{eq:R}) and (\ref{eq:kappae}) \\
    $i$ & orbital inclination & Eq.~(\ref{eq:z})  \\
    $\kappai$ & vertical motion & Eqs.~(\ref{eq:z}) and (\ref{eq:kappai}) \\
    $\chie$, $\chii$ & constant phase angles & Eqs.~(\ref{eq:R}) and (\ref{eq:phi})
    \enddata
\end{deluxetable}

The orbit solutions (Equations (\ref{eq:R})--(\ref{eq:z})) are valid in the limit of small free eccentricity ($\efree \ll 1$) and inclination ($i \ll \pi/2$); also, most parameters are derived using a Taylor-series expansion in terms of $\abin/\rgc$ (e.g., Appendix~\ref{appx:linear}), assuming that the guiding center distance is large compared to the binary semimajor axis. The orbit solutions for a satellite around an eccentric binary contain additional driving terms \citep{leung2013} including one associated with the forced eccentricity, familiar from secular theory \citep[e.g.,][]{murray1999}. This mode, which operates at the frequency $\ngc$, has an amplitude of 
\begin{equation}\label{eq:eforce}
    \eforce \approx  \frac{5}{4}\ebin \frac{\Mp-\Ms}{M}\frac{\abin}{\rgc},
\end{equation}
where $\Mp$ and $\Ms$ are the mass of primary and secondary, respectively, and $M=\Mp+\Ms$.

\subsection{Most-circular orbits}\label{subsec:mostcir}

In the Lee-Peale analytical approximation, orbital motion is the result of a linear superposition of oscillations about a guiding center at harmonics of the synodic frequency and epicyclic motion, characterized by the free eccentricity, $\efree$.  Since a satellite's radial motion has independent contributions from all of these oscillatory modes, trajectories with no free eccentricity tend to stay closer to the guiding center than those for which $\efree > 0$. 

Incidentally, a satellite can be launched on a trajectory with no radial drift, as if it were on a Keplerian circular orbit about the binary center of mass. Mathematically, this condition is met when the contribution from $\efree$ exactly cancels the forced oscillations in Equation~(\ref{eq:R}). However, over time, the epicyclic motion will drive radial excursions that eventually become larger than when $\efree = 0$.  

When a satellite is on a most-circular orbit, with no free eccentricity, it responds to the binary's gravitational influence by drifting slightly away from the barycenter when the secondary is close to it, and shifting inward until the satellite's position is perpendicular to the binary's separation vector.  The extrema in radial distance, which occur whenever $\phi = \phis$ (the satellite is closest to the secondary) and $\phi-\phis = \pm\pi/2$ (the satellite is at right angles to the binary), follow from Equation~(\ref{eq:R}), with $\efree = 0$:
\begin{equation}\label{eq:Rextrema}
  \drplus  =  -\sum_{k=1} C_k \rgc
 \ \ \  \text{and} \ \ \ 
  \drminus  = -\sum_{k=1} C_k \rgc \cos(k\pi/2),
\end{equation}
where $\drpm$ are the extrema in the satellite's radial excursions relative to the guiding center radius, $\rgc$, and the forced mode amplitudes, $C_k$, are from Equation~(\ref{eq:C}). Except for equal-mass binaries, the outward excursion ($\drplus$) is greater in magnitude than the inward excursion ($|\drminus|$).

An order-of-magnitude estimate of the most-circular excursion distances, derived from Equation~(\ref{eq:C}) in Appendix~\ref{appx:linear} using a series expansion with respect to the guiding center distance, is
\begin{equation}
\drpm \sim \rgc \frac{\Mp\Ms}{M^2} \frac{\abin^5}{\rgc^5}
\left(\frac{9}{4} + \frac{3\ngc}{2\nsyn}\right)\left(\frac{\nbin^2}{4\nsyn^2 - \kappae^2}\right). 
\end{equation}
Because the series expansion falls off slowly if a satellite's guiding center distance is only a few times the binary separation, it is prudent to derive values of $\drpm$ directly from Equations~(\ref{eq:Rextrema}) and (\ref{eq:C}), using all available expansion coefficients. 

\begin{figure}[htb]
\centerline{\includegraphics[width=4.5in]{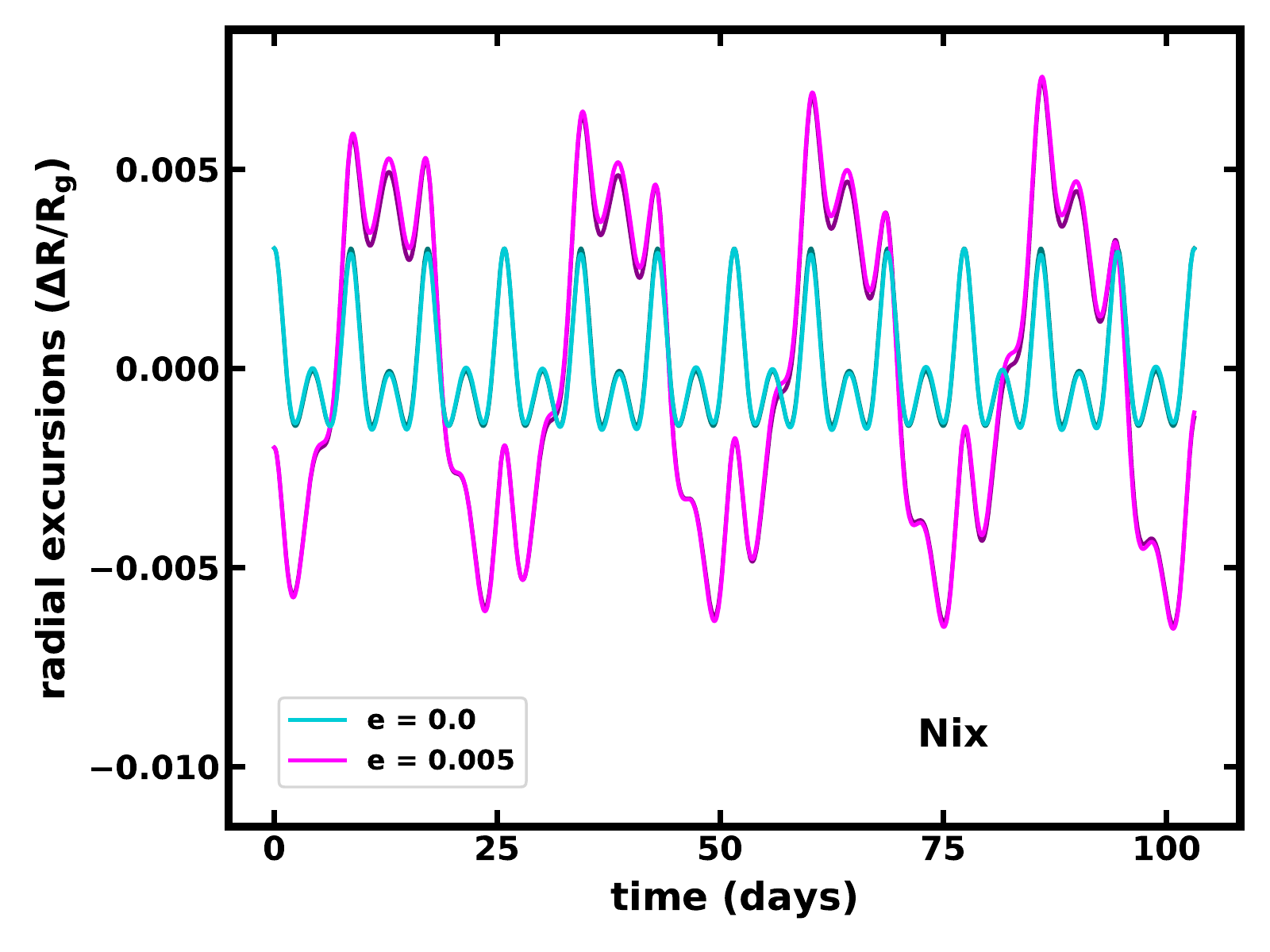}}
\caption{\label{fig:mostcircexample} The radial excursions of Nix on a most-circular orbit (cyan  curve), and on an orbit with free eccentricity $\efree = 0.005$ (magenta curve) are calculated from numerical integration with starting conditions as in Equation~(\ref{eq:R}) at $t=0$. The horizontal axis is time; the vertical axis is the excursion distance, $\Delta R = R-\rgc$, in units of $\rgc$. The analytical predictions for these cases (Eq.~(\ref{eq:R})) are in darker shades. The prediction for the most-circular orbit (dark blue-green) closely tracks its numerically-integrated counterpart (in cyan) and is barely visible behind it. In the mildly eccentric case, the match between the numerical integration and the analytical prediction case (in dark purple, behind the magenta curve) is not as close.}
\end{figure}

Figure~\ref{fig:mostcircexample} provides an illustration of a most-circular orbit based on the small moon Nix around Pluto-Charon. We adopt primary and secondary masses of $1.303\times 10^{25}$ and $1.587\times 10^{24}$~g, and a binary separation of 19,590~km, with Nix at an orbital distance of 2.485$\abin$ \citep{brozovic2015, stern2015, weaver2016, nimmo2017, mckinnon2017}. The figure shows radial excursions relative to the guiding center distance, $\Delta R = R-\rgc$ for Nix on a most-circular orbit ($\efree = 0$) and on one with a small amount of free eccentricity ($\efree = 0.005$). The plots contain the analytical prediction, along with results from a 2-D numerical integration of the moon's equation of motion in the potential of the Pluto-Charon binary, obtained using the Python SciPy routine \texttt{odeint}. 

For orbits with non-zero free eccentricity, the analytical theory is susceptible to errors that scale as $O(\efree^2)$. Thus, it may be necessary to fine-tune starting conditions to derive integrated orbits that match expectations in terms of minimum and maximum radial excursions. To implement this adjustment, we start with the binary and the satellite coaligned and use a 1-D minimization algorithm to vary the azimuthal velocity until long-term orbit integrations yield the desired minimum and maximum excursion distances.

\begin{figure}[htb]
\centerline{\includegraphics[width=4.5in]{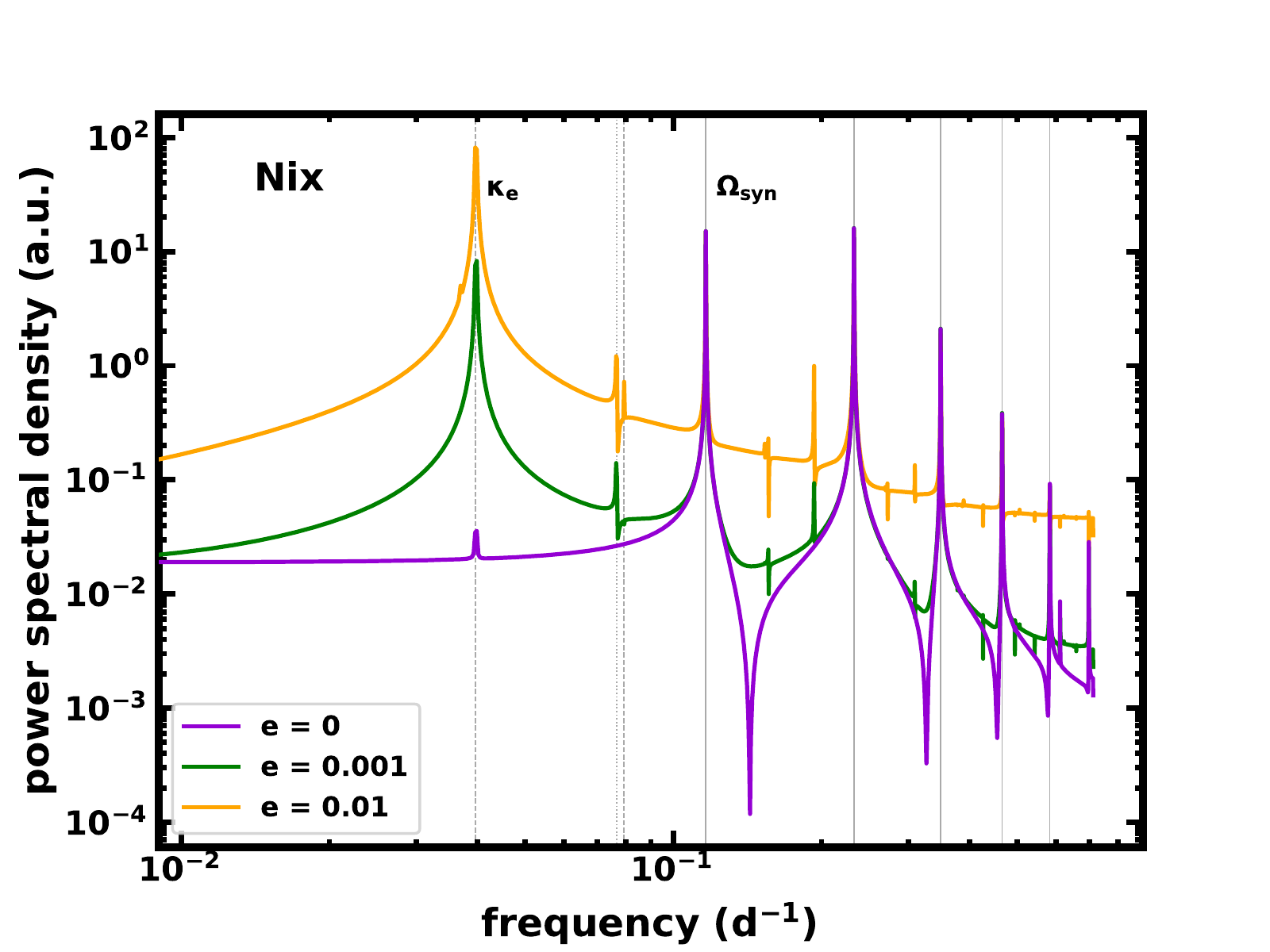}}
\caption{\label{fig:fft} A periodogram of the radial excursions of Nix.  The purple curve shows the power spectral density estimates (in arbitrary units) for the excursions of a satellite on a most-circular orbit. The most prominent peak lies at the synodic frequency $\nsyn$, since the satellite experiences its strongest gravitational kicks at this rate. The vertical gray solid lines are at the locations of $\nsyn$ and its harmonics. There should be no peak at the epicyclic motion $\kappae$ for a most-circular orbit, although there is residual signal from errors in constructing the numerical orbit. The two other curves correspond to orbits with some free eccentricity, as in the legend. In addition to peaks at $\nsyn$ and its harmonics, these curves also have a strong peak at the epicyclic motion, $\kappae$. The gray dashed vertical lines indicate the location of $\kappae$ and its first harmonic, while the dotted line, just to the right of the first harmonic, is a mode ($\nsyn-\kappae$) that is not part of the epicyclic theory. }\end{figure}

In the linearized theory adopted here, the orbital motion is driven at frequencies that include the synodic frequency and its harmonics, and the epicyclic frequency.  Figure~\ref{fig:fft} is a periodogram of the relative radial excursion distance $\Delta R/\rgc$, showing peaks at these frequencies.  The most-circular orbit has only a small residual peak at $\kappae$, indicating no free eccentricity save for numerical noise.  The eccentric orbits ($\efree = 0.01$ and $0.1$) have an additional strong peak at the epicyclic frequency. There are also peaks associated with motion that is not predicted in the analytical model corresponding to terms that would scale non-linearly in $\efree$. These terms, contributing at higher-order harmonics of $\kappae$, as well as $\nsyn-\kappae$, become increasingly important with higher free eccentricity.

\section{Orbit characterization}\label{sec:orbitchar}

Around a single planet or star, standard Keplerian orbital elements completely describe satellite orbits.  The situation is more complicated around a binary, but similar descriptors can quantify orbital properties.  In this section we review a few ways to characterize circumbinary orbits, guided by the results from the linearized theory in \S\ref{sec:theory}.  

\subsection{Keplerian orbital elements}

For orbits around a single central body of mass $M$, the semimajor axis $a$ and the eccentricity $e$ are clearly defined. In terms of a satellite's specific orbital energy $E$ and angular momentum $L$,
\begin{equation}\label{eq:aekep}
\akep = -GM/2E \ \ \ \text{and} \ \ \ \ekep^2 = 1+2|\vec{L}|^2E/G^2M^2.
\end{equation}
Along with $\akep$ and $\ekep$ are the inclination $i$, the epoch $t_0$ (a reference time), the longitude of the ascending node $\ascnode$, the argument of periastron $\omega_p$, and the mean anomaly $M_a$.

For orbits around a binary, neither the orbital energy nor angular momentum of a satellite is conserved.  Nonetheless, when a satellite is far from the binary compared to the binary separation ($\akep \gg \abin$), osculating Keplerian orbital elements are good descriptors of a satellite's trajectory.  Then, the semimajor axis and eccentricity give useful estimates of a satellite's representative orbital distance from the barycenter and the shape of its orbit.

In their analysis of the small moons of Pluto and Charon, \citet{showalter2015} derive Keplerian orbital elements closer to the binary ($\akep \sim 2$--$4\abin$), viewing these elements as describing orbits that are nearly Keplerian when the rapid variations from the binary are averaged out. In their approach, based on secular theory, orbits are elliptical Keplerian trajectories, with some nodal and apsidal precession, superimposed on the rapid forced motion driven by the binary. They fit an elliptical orbit to a set of observations of the position using a nonlinear $\chi^2$ minimization algorithm, expecting residuals to reflect observational errors and the forced motion. For each moon, they derive the six standard Keplerian orbital elements, along with precession rates for the ascending node and the argument of perihelion. This procedure provides a substantially better assessment of Keplerian orbital elements than values determined from individual state vectors and Equation~(\ref{eq:aekep}).

\subsection{Linearized-theory parameters}

Another way to treat the rapid, forced motion of a satellite on a circumbinary orbit is to model it with linearized theory (Eqs.~(\ref{eq:R})--(\ref{eq:z}), for the case of a circular binary). Then, this motion can be isolated from a satellite's Keplerian trajectory, quantified in the theory by the free eccentricity and inclination, along with their associated phase angles. 
The parameters of the theory are\begin{equation}\label{eq:LPparms}
\left\{ t_0, \rgc, \efree, \chie, \inc, \chii ; \Mp, \Ms, \abin, \ebin \right\},
\end{equation}
where the epoch $t_0$ designates the time at which the binary and the satellite's guiding center are aligned (in the case of a circular binary; its value is set to $t=0$ in the preceding section), or when the guiding center is aligned with the binary's argument of perihelion (when $\ebin>0$); all others appear in the orbit solutions, Equations~(\ref{eq:R}) and (\ref{eq:phi}), and in \citet{leung2013}. 

The parameters in Equation~\ref{eq:LPparms} fully describe any orbit within the linearized theory. The model begins to lose accuracy even at modest eccentricity, with errors scaling as $\efree^2$. In simulations of satellite orbits around the Pluto-Charon binary, considered in more detail below, free eccentricities are small \citep[$\efree \sim 10^{-3}$;][]{showalter2015}, and  the formal fractional errors in the orbit solutions are below $10^{-5}$. On the other extreme, Kepler-16b has an eccentricity of about 0.01, but the binary itself has an eccentricity $\ebin = 0.16$; the linearized theory (as extended by \citealt{leung2013}) is first-order accurate in both $\efree$ and $\ebin$, hence orbit-solution errors are at the level of a few percent.
Similarly, the linearized theory is only good to first order in inclination $i$ (Eq.~(\ref{eq:z})).

The approximate correspondence between the linearized theory parameters and the Keplerian ones is
\begin{equation}
  \begin{array}{ccccc}
    \rgc  \rightarrow \akep & \ \ \  & \efree + (\drplus -\drminus)/2\rgc \rightarrow \ekep &
    \ \ \ &  \inc \rightarrow i \\
  \chie \rightarrow -\ascnode-\omega_p & \ \ \ &
  \chii \rightarrow 3\pi/2-\ascnode & \ \ \ & 
  \ngc (t-t_0) \rightarrow M_a
  \end{array}
\end{equation}
provided that a satellite has some low inclination relative to the plane of the binary. Also, both the Lee-Peale parameters and the Keplerian ones are defined relative to the same reference direction, chosen here to be the secondary's position vector at epoch $t_0$.

\subsection{Geometric parameters}

A satellite's semimajor axis and eccentricity, which are defining properties of a Keplerian orbit, may be generalized to characterize bound orbits in a variety of systems, including galaxies \citep[e.g.][]{lyndenbell1963}. In these broader contexts, their purpose is to provide the typical distance from the system barycenter and the extent of the radial traverses relative to a dynamically cold --- possibly circular --- orbit.  For stable circumbinary orbits, we might choose geometric measures based on the extrema of a satellite's radial excursions;  these quantities contain information about a satellite's motion on rapid time scales, which is missed by secular perturbation theory. They also give useful information when a satellite's orbit has radial excursions that are too large to be accurately described by linearized theory (e.g., $\Delta R/\rgc \gtrsim 0.1$).

Following this strategy, we propose geometric-based orbital elements that reflect the extrema of a satellite's radial excursions, but are also corrected for forced motion caused by the binary:
\begin{eqnarray}\label{eq:ageo}
  \ageo & = & \frac{1}{2} \left[(\rmax + \rmin) - (\drplus + \drminus)\right]
  \\
  \label{eq:egeo}
  \egeo & = & \frac{1}{2}\left[(\rmax-\rmin)-(\drplus-\drminus)\right]/\ageo\!,
\end{eqnarray}
where $\rmin$ and $\rmax$ are the minimum and maximum values of the satellite's radial coordinate over some long time interval, and $\drpm$ are the extrema of a dynamically cold orbit at this orbital distance. The second term in the right-hand side of each definition is the correction for the forced motion, which is not symmetric about mean orbital distance ($\drplus > \drminus$). Because an estimate of the forced motion requires knowledge of the orbital distance, we adopt Equation~(\ref{eq:Rextrema}), evaluated at $\rgc = (\rmax-\rmin)/2$. Then we use iterative improvement, accomplished by setting $\rgc = \ageo$ in the analytical expressions for $\drpm$, which converges quickly.

Our implementation of these geometric measures relies on the linearized theory to quantify the forced motion. However, the theory also only applies for small eccentricities, $\efree, \ebin \ll 1$. For example, when $\ebin \gtrsim 0.1$, we estimate relative errors in analytically-derived $\drpm$ to be $\gtrsim 10^{-3}$. Numerical orbit integrations, used to make this error estimate, can assist in more accurately finding the true radial extent of dynamically cold orbits. These corrections become less important when the satellite's periastron distance from the barycenter is large compared with the binary separation ($(1-\egeo)\ageo \gg \abin$ ). Then, $\ageo$ and $\egeo$ give good approximations to their Keplerian counterparts.

Next we consider how to extract the parameters described here from numerical simulations or observational data.

\subsection{Parameter estimation}

An ideal set of observations of a circumbinary moon or planet, sampled densely over a long period of time, yields any desired set of orbital elements.  As in \citet{woo2020}, even just a single snapshot of a satellite and its binary host can produce such data, if observed positions and velocities are numerically integrated in time. Fitting these integrated orbits to other numerically generated ones with known properties is sufficient to estimate any set of orbital parameters. If the Lee-Peale theory applies, the problem is reduced to one of model fitting in a three- or five-dimensional parameter space, depending on whether the binary is eccentric. \citet{woo2018,woo2020} provide an FFT-based alternative to this approach.

By comparing observed orbits to an exhaustive suite of numerical or theoretical ones with known characteristics, the problem of parameter estimation is solved. However, the fitting process is computationally slow relative to calculating osculating Keplerian orbital elements; both the semimajor axis and eccentricity derive from a measurement of energy and angular momentum at a single epoch (Eq.~(\ref{eq:aekep})). The issue of computational resources becomes a problem in large-scale simulations of many particles in a circumbinary environment, where estimation by detailed orbit fitting is not feasible. However, since simulations are designed to track particle orbits, with a small expenditure of memory and little computational overhead we can estimate the geometric parameters $\ageo$ and $\egeo$ (Eqs.~(\ref{eq:ageo}) and (\ref{eq:egeo})) without any orbit fitting. 

Here we offer another option, a set of orbital parameter estimators derived from single snapshots of a satellite's position and velocity. The starting point is linearized theory and the free eccentricity:
\begin{equation}\label{eq:eccest}
  \eest \cos(\chie) = 
  \frac{\rddobs-\rddtheory}{\kappae^2 \rgc}
    \ \ \ \ \ 
  \eest\sin(\chie) = -\frac{\phiddobs-\phiddtheory}{2\kappae\ngc}
\end{equation}
where $\eest$ denotes an estimated quantity, and $\robs$ and $\phiobs$ are the observed cylindrical coordinates of a satellite; $\rtheory$ and $\phitheory$ are theoretical expectations for a most-circular orbit (Eqs.~(\ref{eq:R}) and (\ref{eq:phi}) with $\efree = 0$). We approximate the theoretical quantities, including $\ngc$ and $\kappa$, by adopting a guiding center radius $\rgc = \robs$. We also need the time variable to use in the formulae for the theoretical coordinates, which we glean from the angular coordinate of the satellite $\phiobs$ and an estimate of its angular speed $\phidotobs$. For example, in the case of a circular binary,  we need the time since the binary and the satellite were most recently coaligned, which is approximately $(\phis-\phiobs)/\phidotobs$. A similar prescription applies for an eccentric binary \citep{leung2013}, where the relevant quantity is the time since the satellite was positioned at the binary's argument of periastron.

Our general strategy for selecting the free eccentricity estimator, $\eest$ in Equation~(\ref{eq:eccest}), is to choose some observable quantities based on radial and azimuthal coordinates, and then subtract off the contributions that are associated with most-circular orbits to isolate the contributions from the free eccentricity. For example, $\robs - \rtheory \sim \rgc \eest \cos(\chie)$, and $\phiobs-\phitheory \sim \ngc\eest\sin(\chie)$, are possibilities. However, the theoretical values contain quantities that are not always well-constrained; the leading term in $\rtheory$ is $\rgc$, which carries some uncertainty that can dominate the difference $\robs-\rtheory$. We therefore adopt the second-order time derivatives in Equation~(\ref{eq:eccest}), which do not depend on uncertain constant terms. 

\begin{figure}[htb]
  \centerline{
    {\includegraphics[width=3.5in]{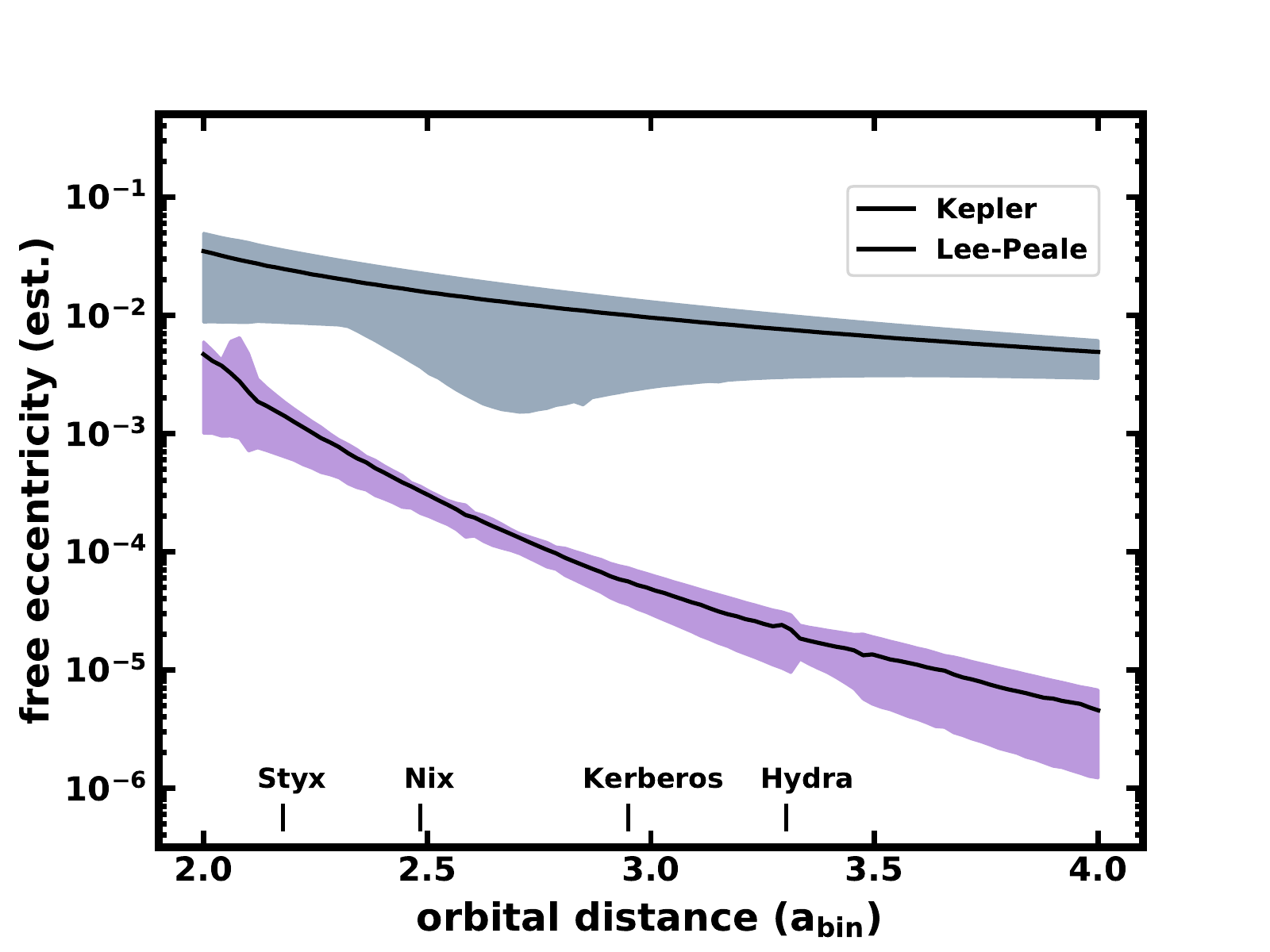}}
    {\includegraphics[width=3.5in]{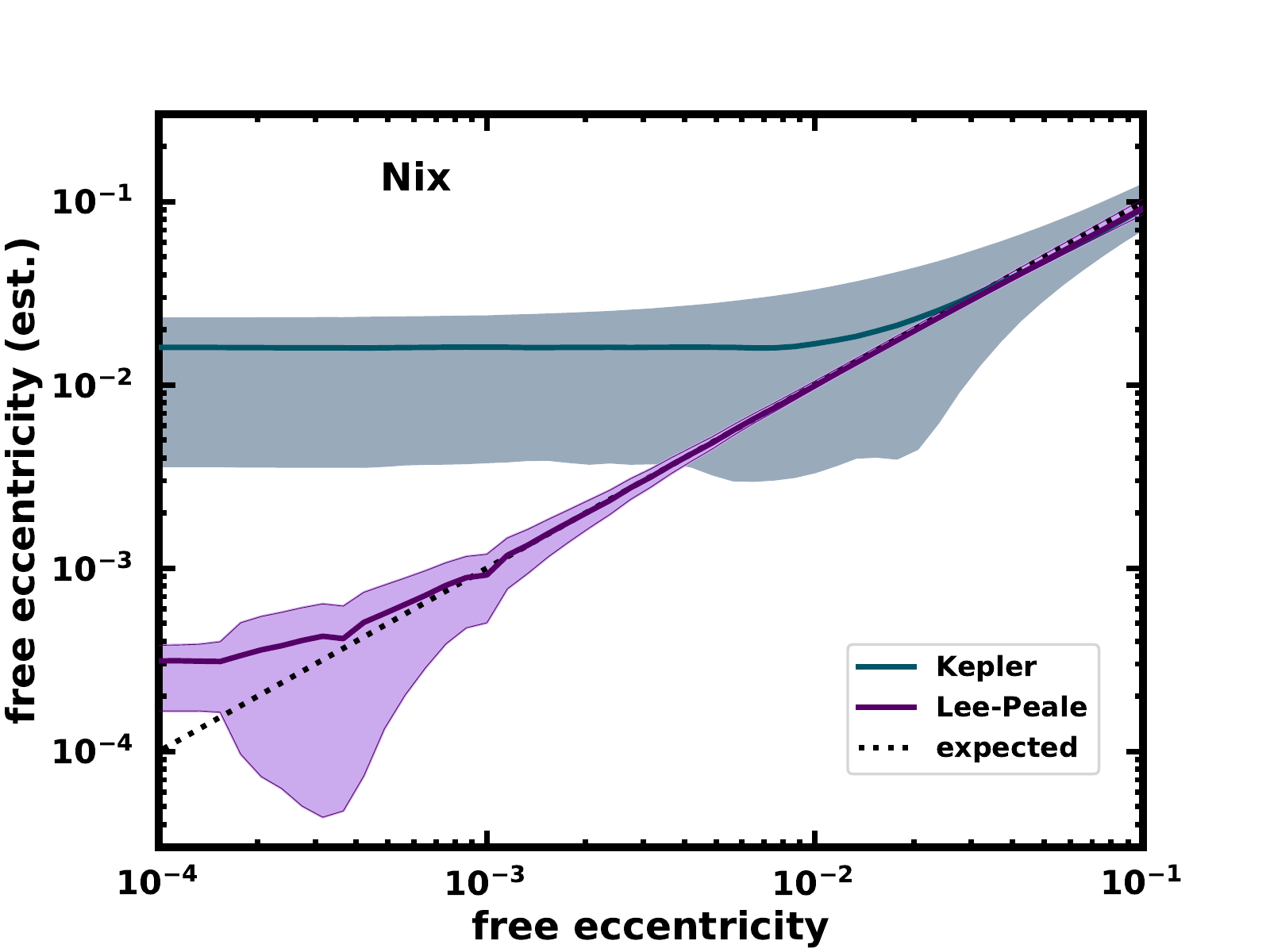}}
  }
  \caption{\label{fig:eccvs} 
  Estimation of free eccentricity for simulated orbits around the Pluto-Charon binary based on $\ekep$, the osculating Keplerian eccentricity  (Eq.~(\ref{eq:aekep}); green colors) and $\eest$, the estimator from the linearized approximation (Eq.~(\ref{eq:eccest}); purple colors). The shaded areas show the central 95th-percentile range of samples taken during orbit integrations, and the dark curves are median values.  The left plot shows most-circular orbits, with an expected free eccentricity of zero, as a function of the orbital distance.  The locations of the binary planet's moons are indicated for reference.  The Keplerian estimator $\ekep$ cannot resolve $\efree$ values below $\sim 10^{-2}$, while $\eest$, from linearized theory, achieves significantly better resolution.  The right-hand plot shows the estimated free eccentricity as a function of the expected value (dotted line) for orbits at the distance of the moon Nix ($\rgc = 2.485\,\abin$). The various undulations in the shaded regions may be attributed to resonances (e.g., near the orbit of Styx in the left-hand plot) and finite sampling. 
    }
\end{figure}

Using orbits around the Pluto-Charon binary as an example, we show the effectiveness of $\eest$ (Eq.~(\ref{eq:eccest})) and $\ekep$ (Eq.~(\ref{eq:aekep})) in Figure~\ref{fig:eccvs}. The curves in the figure reveal general trends in the performance of these estimators, while the shaded regions illustrate that the estimators give a range of values from single-epoch measurements taken at various orbital phases. Both estimates are more precise at larger orbital distance, and each has a lower limit to the free eccentricity that it can resolve. When applied to most-circular trajectories ($\efree = 0$), the linearized-theory estimator gives values of $\efree$ within $0.01$ for orbits close to the binary ($\rgc \sim 2\abin$) and below $10^{-5}$ at orbital distances more than a few time the binary separation. This performance compares well with the values of eccentricity derived from osculating Keplerian orbits. The Keplerian measure
in Equation~(\ref{eq:aekep})) is not designed to account for the effect of the binary, and thus cannot resolve eccentricities much below $0.01$ for the range of orbital distances or expected free eccentricities shown in the figure. 

\begin{figure}[htb]
  \centerline{
    {\includegraphics[width=3.5in]{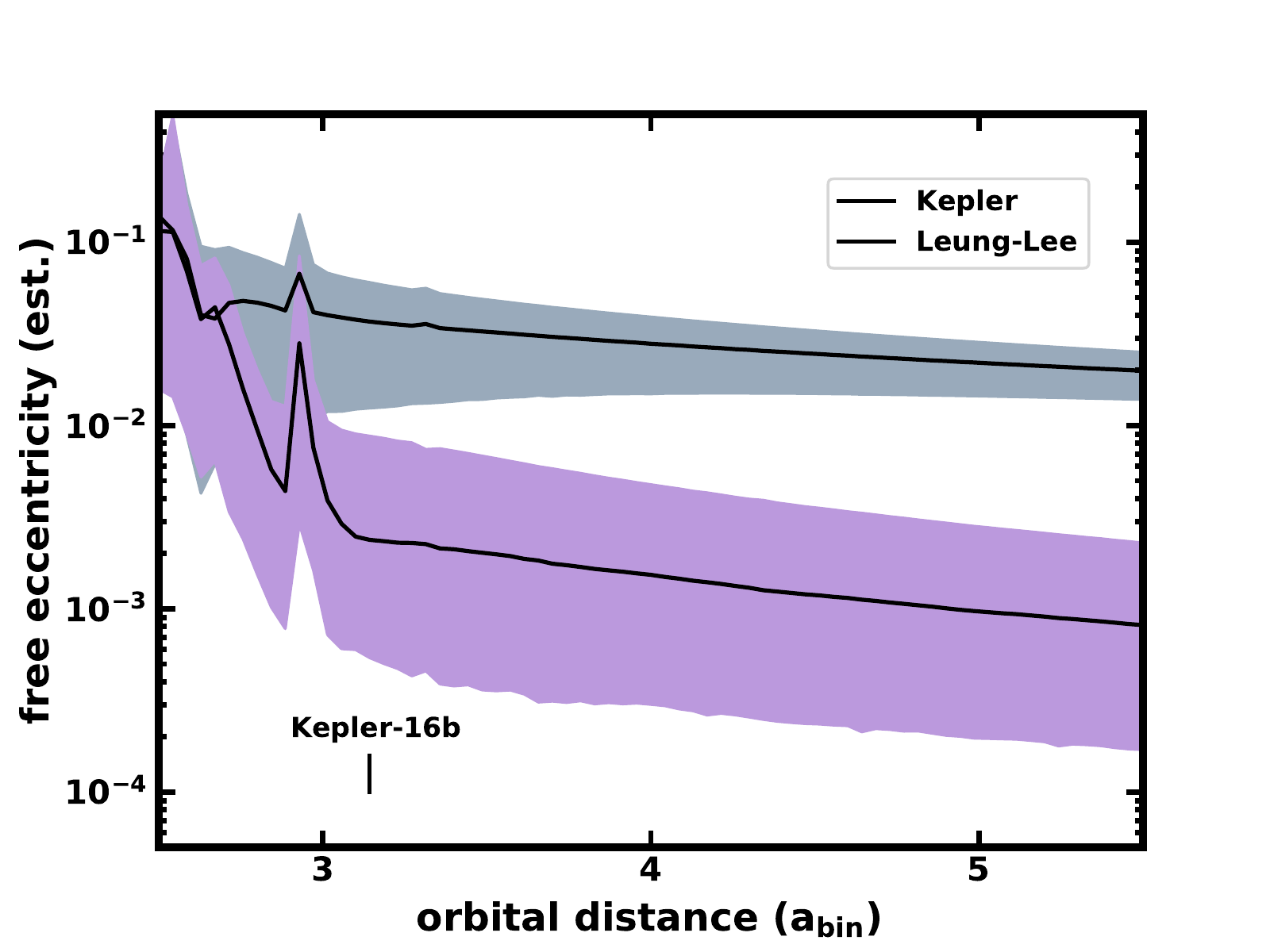}}
    {\includegraphics[width=3.5in]{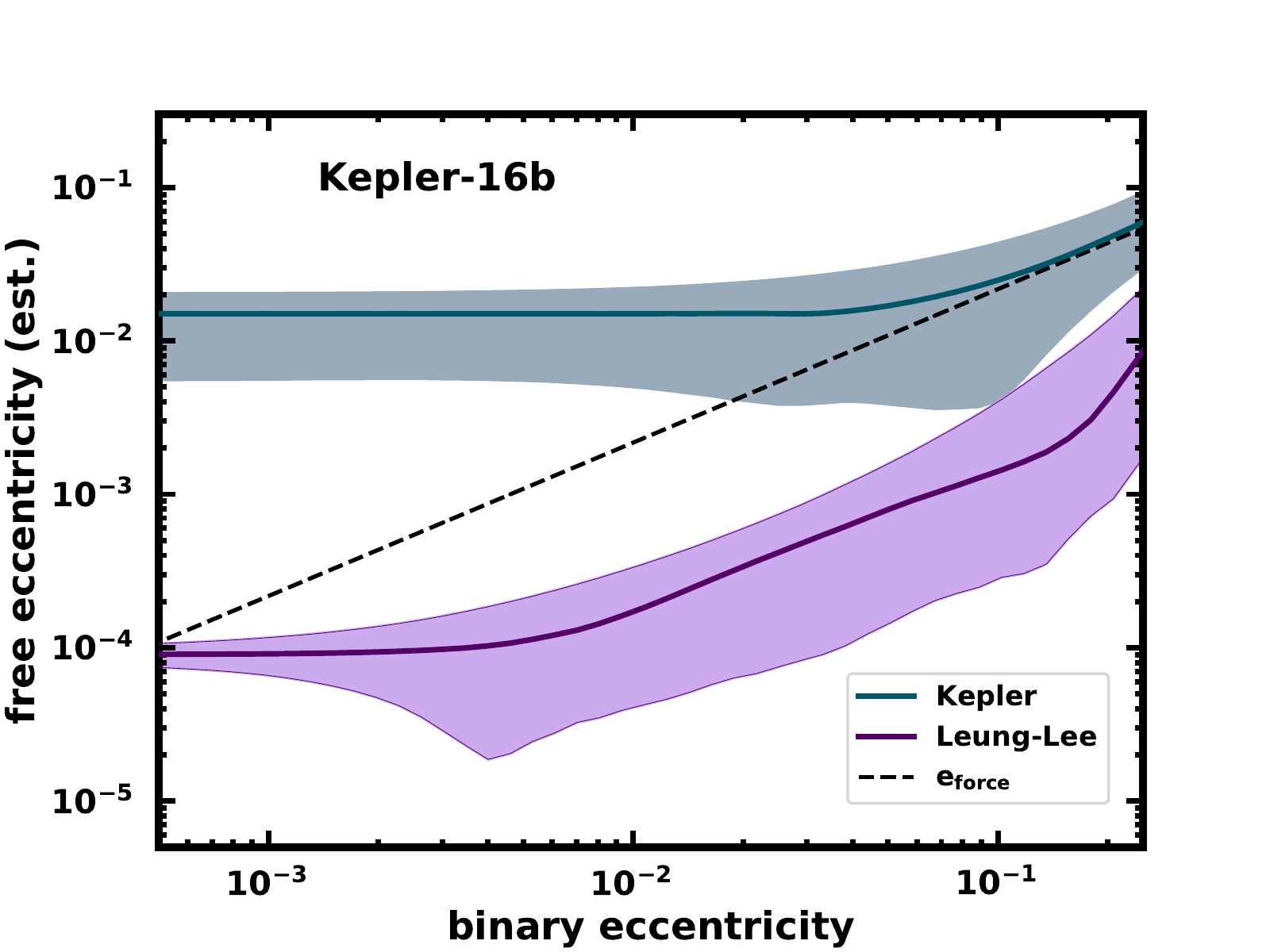}}
  }
  \caption{\label{fig:eccbin} 
  Estimates of free eccentricity using linearized-theory and osculating Keplerian measures for most-circular orbits around an eccentric binary based on Kepler-16. As in Figure~\ref{fig:eccvs}, the left plot shows $\ekep$ (Eq.~(\ref{eq:aekep})) and $\eest$ (Eq.~(\ref{eq:eccest})) as a function of the orbital distance when the expected $\efree$ is zero and the binary's eccentricity is $\ebin = 0.16$.  The Keplerian estimates include the forced eccentricity (Eq.~(\ref{eq:eforce})), which has a value of about 0.03 at the distance of the planet Kepler-16b. The linearized theory estimator can resolve free eccentricity below this value.  The undulations in both estimators stem from finite sampling and resonances, including the 5:1 mean-motion resonance at 2.9$\abin$. The right plot shows how the estimators depend on the binary eccentricity.  For reference, the amplitude of the forced eccentricity is included (dashed line). The values of $\ekep$, which do not distinguish free and forced modes, resolve $\eforce$ at high $\ebin$. The linearized theory estimator tracks only $\efree$, resolving it to within $\sim 10^{-4}$ at low $\ebin$. However, its errors, which scale as $\ebin^2$, grow with increasing binary eccentricity.
  }
\end{figure}

Figure~\ref{fig:eccbin} illustrates how eccentricity estimators perform for most-circular orbits ($\efree = 0$) when the central binary itself has some eccentricity.  The plots in the figure are based on the Kepler-16 system, with primary and secondary masses $0.6897$~\Msolar\ and $0.20255$~\Msolar, binary separation $\abin = 0.2243$~au, and eccentricity $\ebin = 0.16$ \citep{doyle2011}. The circumbinary planet Kepler-16b is at an orbital distance of 0.7048~au from the system barycenter.  As the plots illustrate, the performance of $\eest$ degrades as the binary eccentricity increases, since the underlying theory is linear in binary eccentricity as well as in $\efree$ \citep[see][]{leung2013}. The peaks in the curves at small orbital distance (left panel) are the result of resonances; the largest peak is the 5:1 commensurability. The upper limit of binary eccentricities shown in the figure (right panel) is about 0.3, as orbits at Kepler-16b's distance become unstable when $\ebin$ rises above this value \citep{popova2013, chavez2015}.

Despite the challenges for the linearized theory with eccentric binaries, the results in Figure~\ref{fig:eccbin} are promising. The eccentricity estimator $\eest$ can resolve $\efree$ at the level of a few time $10^{-3}$ for orbits at Kepler-16b's location, even when $\ebin \approx 0.16$. 

\begin{figure}[htb]
  \centerline{
    {\includegraphics[width=5.5in]{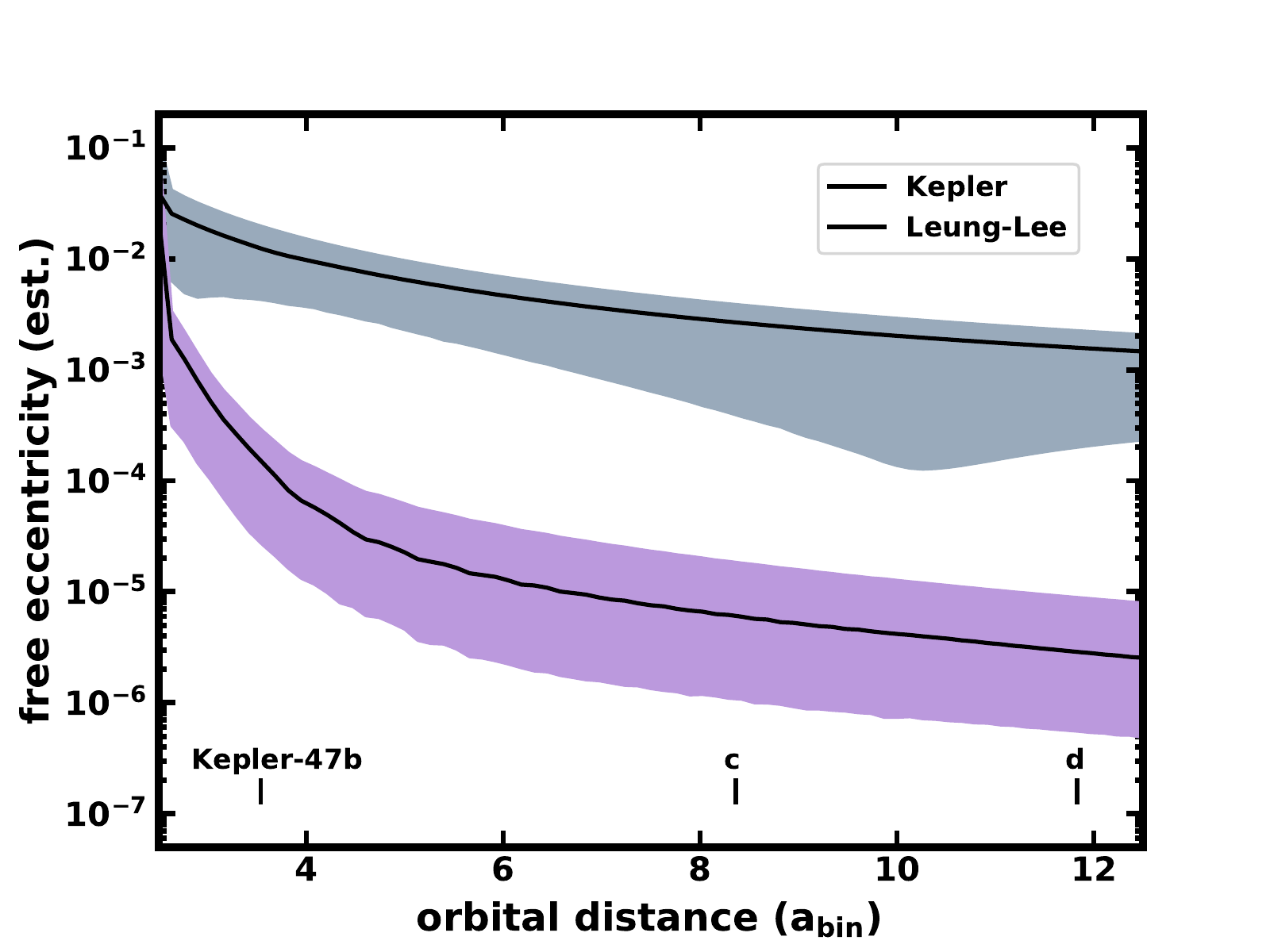}}
  }
  \caption{\label{fig:eccbinkep47} Estimates of free eccentricity using linearized-theory and osculating Keplerian measures for simulated most-circular orbits around Kepler-47, similar to the left panel in Fig.~\ref{fig:eccbin}.  Because of Kepler-47's smaller binary eccentricity, the Leung-Lee \citeyearpar{leung2013}) linearized theory yields an estimator that is more sensitive to free eccentricity than for orbits around Kepler-16.}
\end{figure}

We expect better performance with the estimator $\eest$ in circumbinary systems like Kepler-47, whose central binary is much less eccentric \citep[$\ebin=0.0234$; $M_p = 1.043$~\Msolar, $M_s = 0.362$~\Msolar, $\abin = 0.0836$~au][]{orosz2012a, kostov2013}. Figure~\ref{fig:eccbinkep47} provides an example, showing that the free eccentricity of the three Kepler-47 planets can be resolved down to levels of $10^{-4}$ or less.

Our focus thus far has been on fast, single-epoch estimators of eccentricity. We turn now to a separate estimator for orbital distance, applicable in the limit of small binary eccentricity, $\ebin \lesssim 0.1$. While the geometric estimator $\ageo$ is robust and may be accumulated with little cost to computational load, we explore how to estimate orbital distance from a single state vector.  The starting point is the Jacobi integral, which is a constant of motion for planets or moons around circular binaries: 
\begin{eqnarray}\label{eq:jacobi}
  \Cj & \equiv & 2\nbin L - 2E
  \ \ \ \ \ \ \ \ \ \ \text{(circular binary, coplanar orbit)},
\end{eqnarray}
which we connect to orbital elements using Equation~(\ref{eq:aekep}). In the limit of zero free eccentricity, the angular momentum is $L \approx \ngc\rgc^2$, while the energy is $E \approx L^2/2\rgc^2 + \Phi_0$,  with $\Phi_0$ defined as the orbit-averaged potential of the central mass (Eq.~(\ref{eq:Phik})). By including an eccentricity-dependent factor that accounts for the reduction in a satellite's angular momentum as the eccentricity increases, we have  
\begin{eqnarray}  
  \label{eq:jacobimostcirc}
\Cj & \approx & (1-\efree^2)^{1/2}(2\nbin-\ngc)\ngc\rgc^2 - 2\Phi_0   \ \ \ (e \ll 1).
\end{eqnarray}
 To obtain an estimate of the guiding center distance, $\rgcest$,  from $\Cj$, we assume a value for $\ngc$, based on the observed radial position and/or angular speed, and set $\efree^2$ to zero in accordance with Lee-Peale theory. Then. we recalculate $\ngc$ using the estimate $\rgcest$ in Equation~(\ref{eq:ngc}), and iterate to a desired precision. 

The orbital distance measure $\rgcest$, obtained from iteration of  Equation~(\ref{eq:jacobimostcirc}), can be extended to accommodate non-zero free eccentricity. In a ``hybrid'' approach that includes the factor ($1-\efree^2$) within the angular momentum term in Equation~(\ref{eq:jacobimostcirc}), estimates of $\efree$ from Equation~(\ref{eq:eccest}) enter quadratically into the iteration.

\begin{figure}[htb]
{ \raisebox{0.1in}{\includegraphics[width=3.5in]{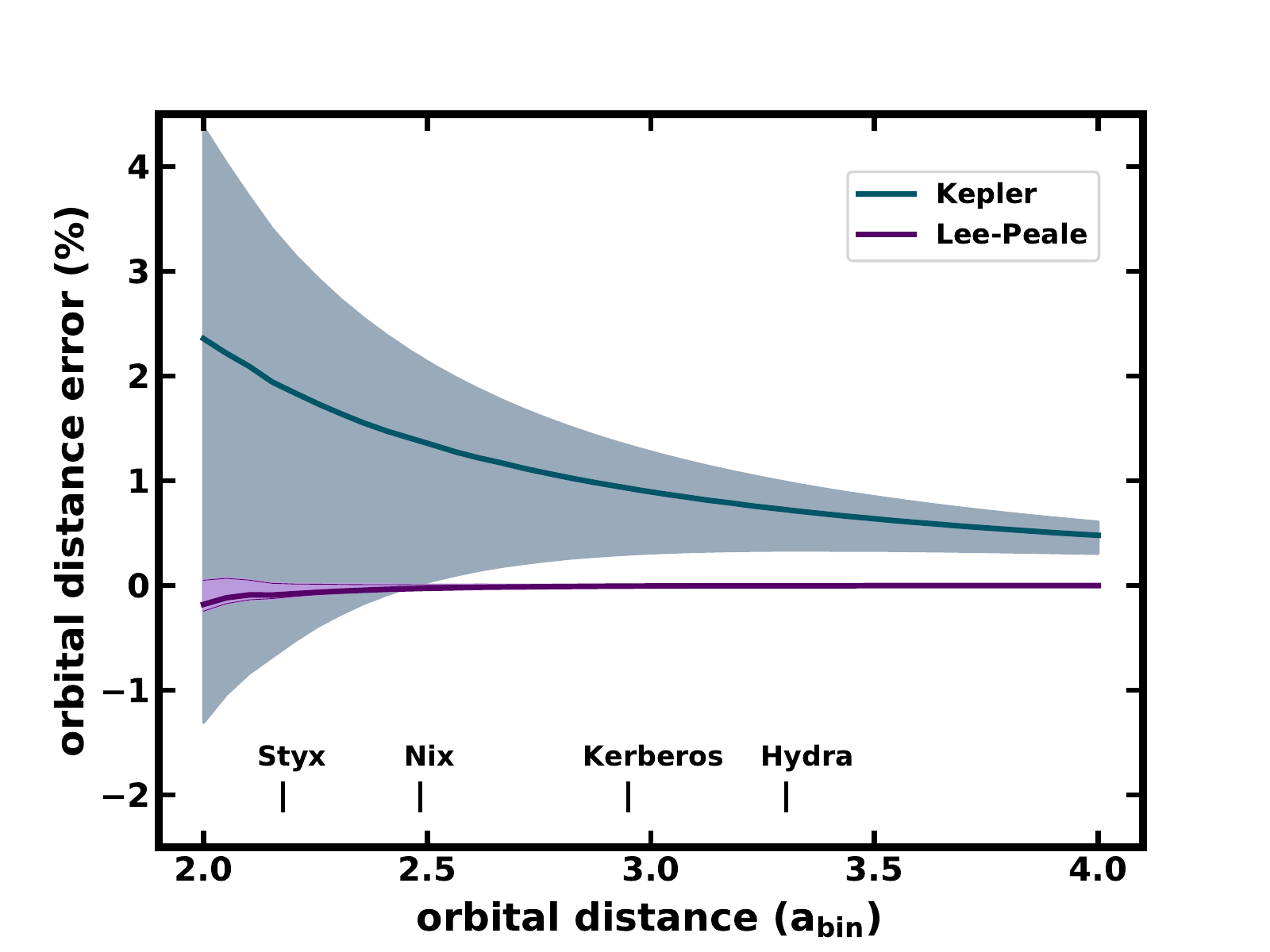}}
\includegraphics[width=3.4in]{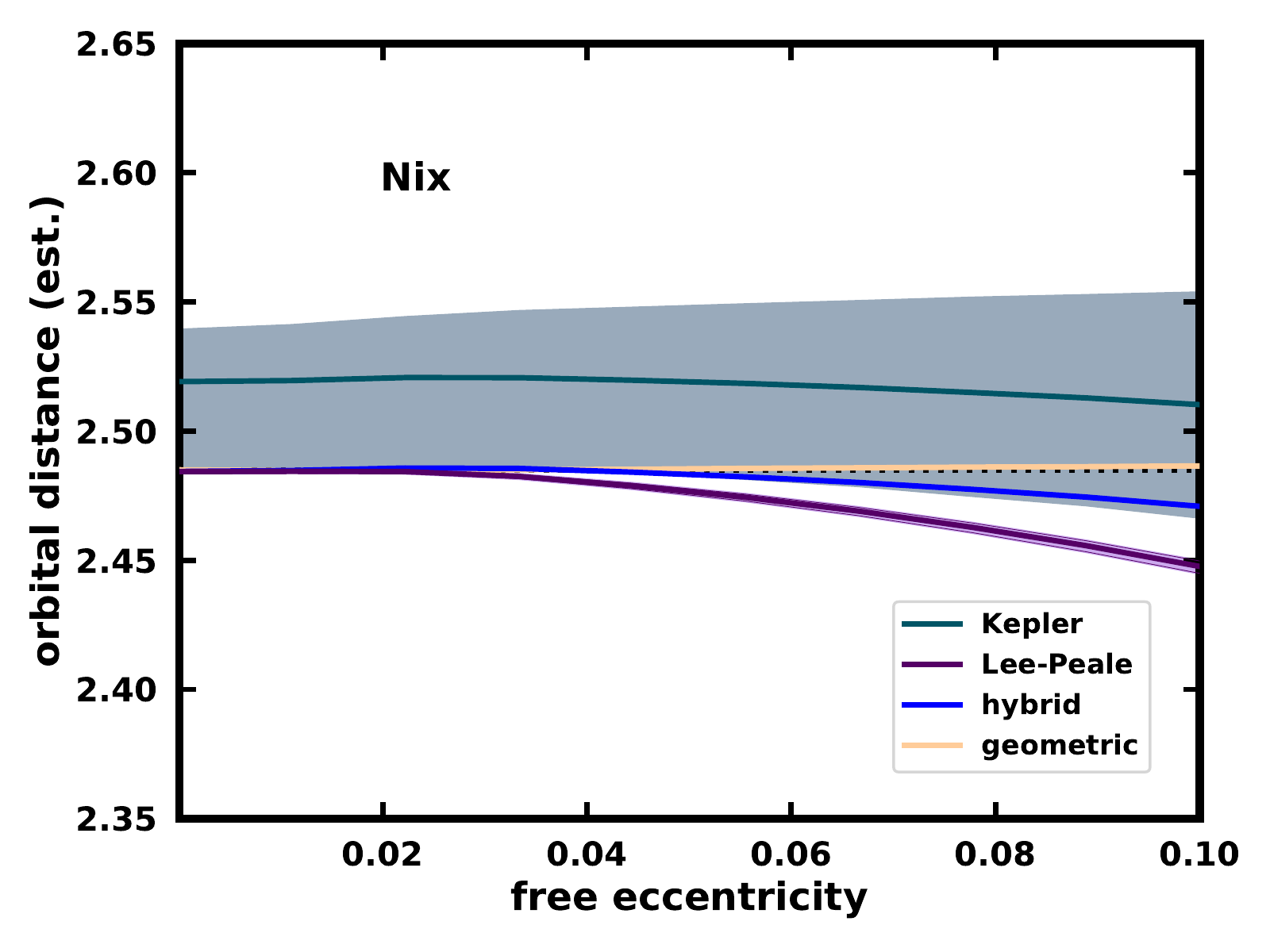}}
\caption{\label{fig:rgcvs} 
   Estimators of a satellite's average orbital distance from simulations of orbits around the Pluto-Charon. The left panel shows the error in distance estimates $\rgcest$ (from linearized theory, Eq.~(\ref{eq:jacobimostcirc}), in purple) and $\akep$ (the osculating semimajor axis, Eq.~(\ref{eq:aekep}), in green)  for most-circular orbits at various orbital distances. The error is given as the percent difference between the estimate and the expected value. The shaded regions show the central 95\% of measurements in snapshots from orbit integrations.
   The right panel shows these estimators along with the ``hybrid'' $\rgcest$ described in the text, and the geometric estimator (Eq.~(\ref{eq:ageo})) for satellites with a range of free eccentricity at the orbital distance of Nix ($\rgc = 2.485\,\abin$; the dotted black line). 
}
\end{figure}

In Figure \ref{fig:rgcvs}, simulated orbits around the Pluto-Charon binary illustrate the application of orbital distance estimates --- the osculating semimajor axis $\akep$, the geometric measure $\ageo$, the Jacobi-integral measure $\rgcest$ with $\efree = 0$, and its hybrid form that includes eccentricity estimates. The figure shows the error in $\akep$ and in $\rgcest$ relative to the expected value (e.g., $(\rgcest-\rgc)/\rgc$) in integrations of most-circular orbits. The Jacobi-integral estimator, $\rgcest$, is accurate to within a fraction of a percent in the vicinity  of the Pluto-Charon moons, while the osculating semimajor axis is different from the expected orbital distance by a percent or larger in this same region. The figure also demonstrates that as the free eccentricity increases, the hybrid version of $\rgcest$ is preferable when only a single snapshot of an orbit is available. Otherwise, if samples of radial positions are available over many orbits, we strongly recommend the geometric estimator, $\ageo$.

Numerical results of the orbital distance measure derived from Equation~(\ref{eq:jacobimostcirc}) with eccentric binaries suggests that errors in estimated $\rgc$ are below 0.2\%\ (95\% confidence) if the binary eccentricity is below $\ebin \approx 0.05$ for orbits near 3$\abin$. For Kepler-16b, the errors are roughly 0.5\%. The osculating Keplerian semimajor axis, if used as a surrogate for circumbinary orbital distance, overestimates the guiding-center distance by about 1\%.

To summarize this section, Table~\ref{tab:estimators} lists the estimators considered here to characterize circumbinary orbits. Our next step is to apply these estimates to data, both real and simulated.

\begin{deluxetable}{llll}
  \tablecaption{\label{tab:estimators} 
  Estimators to characterize circumbinary orbits}
  \tablehead{ \colhead{parameter} & \colhead{description} &
  \colhead{applicability} & \colhead{reference}}
    \startdata
    \multicolumn{4}{c}{Keplerian} \\
    \tableline
    $\akep$ & osculating semimajor axis \ \ \  & single snap-shot & Eq.~(\ref{eq:aekep}) \\
    $\ekep$ & osculating eccentricity & $R \gg \abin$; $\efree \gtrsim 0.01$ \ \ \ \ \ \ \ \ \ \ \ \ &  Eq.~(\ref{eq:aekep})
    \\
    \tableline
    \multicolumn{4}{c}{Geometric}
    \\
    \tableline
    $\ageo$ & orbital distance & many samples & Eq.~(\ref{eq:ageo}) 
    \\
    $\egeo$ & free eccentricity & any stable orbit & Eq.~(\ref{eq:egeo})
    \\
    \tableline
    \multicolumn{4}{c}{Linearized theory}
    \\
    \tableline
    $\rgcest$ & guiding center distance & single snapshot & See Eq.~(\ref{eq:jacobimostcirc}) 
    \\
    $\eest$ & free eccentricity & $R\gtrsim 3\abin$; $\efree$, $\ebin \lesssim 0.1$ & Eq.~(\ref{eq:eccest})
    \enddata
\end{deluxetable}

\section{Application to observations: a comparison}\label{sec:comp}

In this section, we report on estimation of the eccentricity of the small moons of Pluto-Charon, with the goal of comparing the linearized-theory prescription (Eq.~(\ref{eq:eccest})) and the geometric measure (Eq.~(\ref{eq:egeo})) with published results.  

The Pluto-Charon system is nearly co-planar, with all members on low-eccentricity orbits. Thus its orbital dynamics lie solidly in the domain of the Lee-Peale linearized theory.  Using the single-epoch state vectors provided by \citet[see Table~8  therein]{brozovic2015}, along with the binary masses, we estimate the instantaneous free eccentricity for each of the small moons from Equation~(\ref{eq:eccest}), deriving accelerations (e.g., $\rddobs$), from Newton's Law of Gravity. To get the geometric free eccentricity estimate $\egeo$ (Eq.~(\ref{eq:egeo})), we integrate the \citet{brozovic2015} state vectors forward in time, sampling radial locations throughout. Table~\ref{tab:ecc} list the derived values, the FFT-based measures from \citet{woo2018, woo2020}, and results from the orbit fits to a Keplerian ellipsoidal trajectory from \citet{showalter2015}.

\begin{deluxetable}{lcccc}
  \tablecaption{Estimates of the free eccentricity of Pluto-Charon's moons}
  \tablehead{ \colhead{Name} & \colhead{~~~~~~linearized~~~~~~} &
    \colhead{~~~~~~geometric~~~~~~} & \colhead{~~~~~~~FFT~~~~~~~} & \colhead{~~~~~~orbit fit~~~~~~}}
    \startdata
    Styx & 0.00299 & 0.00127 & 0.00110 & 0.00579 \\
    Nix & 0.00213 & 0.00192 & 0.00187 & 0.00204 \\
    Kerberos & 0.00327 & 0.00376 & 0.00320 & 0.00328 \\
    Hydra & 0.00559 & 0.00561 & 0.00551 & 0.00586 \\
    \enddata
  \tablecomments{\label{tab:ecc} The linearized-theory result
    (Eq.~(\ref{eq:eccest})), is applied to the single-epoch,
    phase-space data from \citet[Table 8 therein]{brozovic2015}.  The
    geometric estimate is from Eq.~(\ref{eq:egeo}) and a numerical
    integration of the same data. The FFT-based estimate is from
    \citet{woo2020}, while the last column is from a fit to an
    ellipsoidal Keplerian trajectory \citet{showalter2015}.}
\end{deluxetable}

Table~\ref{tab:ecc} shows strong similarities between the various measures of free eccentricity. The biggest differences arise for Styx, which is at an orbital distance of about $2.2 \abin$. This small distance is a challenge for linearized theory. The value of $\eest$ from the \citet{brozovic2015} state vector is $\eest = 0.00299$; estimates applied to a numerical integration of that same vector yields $\eest = 0.00028$--$0.00319$ (including 100\%\ of the samples). Measuring $\eest$ from samples of a numerically integrated most-circular orbit at Styx's location yields values of $\eest$ no higher than $0.00215$. Thus,  we confirm with full confidence that Styx has some free eccentricity. 

The variation of free eccentricity measurements of the other moons in numerical integrations of the \citet{brozovic2015} state vector are significantly smaller (well below $10^{-3}$). The values of $\eest$ for these outer moons in Table~\ref{tab:ecc} are also more consistent with estimates of the free eccentricity from \citet{woo2020} and \citet{showalter2015}.

We conclude that the single-epoch estimator of eccentricity in Equation~(\ref{eq:eccest}) is competitive with other estimators, at the level of measuring small values of $\efree \sim 10^{-3}$. We take advantage of this result next.

\section{Application in numerical simulations}\label{sec:apps}

Here, we apply the linearized-theory estimator $\eest$ in a simulation of eccentricity damping in a circumbinary environment. This section presents the main result of this work, a demonstration that we can quickly track and modify orbital elements in a simulation with $n$-bodies on circumbinary orbits. We begin with some context from planet formation theory.

Planets or satellites grow from small particles in orbit around a central body because of collisions. Dust grains in protoplanetary disks form into cm-size pebbles \citep[e.g.,][]{vandermarel2015, birnstiel2016} and then into planetesimals \citep[][and references therein]{youdin2002, yk2013, johansen2014, simon2017}, which in turn grow by accretion and mergers into protoplanets whose gravity helps to accumulate the smaller solids around them \citep{gold2004, kb2009, ormel2010a, bk2011a, lamb2012, chambers2014, levison2015, johansen2015, eriksson2020, morbidelli2020}. Collisional damping and dynamical friction reduce relative speeds, facilitating mergers and growth. However, the large protoplanets gravitationally stir smaller bodies to high speeds, triggering a collisional cascade of destructive collisions, grinding small solids into dust that is removed by stellar wind or radiation. The balance between stirring and damping, which governs how planets emerge, hinges critically on the outcome of collisions throughout this process.

Collision outcomes depend on the relative speeds between bodies and their bulk strength \citep[e.g.,][]{benz1999, lein2009}. Material strength makes very small particles hard to break --- rocky material is harder to shatter than ice ---  while self-gravity holds together the very large ones. The weakest bodies are intermediate-sized, with radii of 1--10~km; collision speeds as low as 10-100~m/s are sufficient to disrupt them.  In simulations of planet or satellite formation, we must be able to track mutual collision speeds at least slow as these values. 

In a swarm of particles round a single central body, the Keplerian eccentricity is an excellent measure of pairwise collision speeds. The idea is that as damping dynamically cools the swarm, they tend toward nested, circular orbits that do not cross. Collisions occur only when particles have random motion relative to these circular paths. Their random speeds are $\vrnd \sim e \vkep \sim e\akep\nkep$, where $e$ is the eccentricity and $\vkep$ is the speed of a particle on a circular orbit. Hence, eccentricity is an indicator of collision outcomes. As an example, for particles at 1~au around the Sun, the minimum disruption speed for the weakest rocky bodies, $\sim0.1$~km/s, corresponds to $e \approx 0.003$. When eccentricities damp below this value, all collisions lead to coagulation and growth.

Around a central binary, dynamical cooling through orbital damping operates much the same way, with particles settling onto most-circular orbits. These nested, non-crossing trajectories are the analogs to circular orbits around a single star. In the circumbinary case, random motion is associated with the free eccentricity, as in Equation~(\ref{eq:R}),  with $\vrnd \sim \efree\rgc\ngc$. Because the random speeds are important in determining the outcome of interactions between solid particles on orbits that may cross, monitoring $\efree$ is critical to tracking the velocity evolution of a circumbinary particle disk. Once interactions have been identified, we may modify random motions to reflect the physics of gravitational stirring, damping, and/or disruptive collisions. 

Simulations of planet formation thus follow the eccentricity of solids as their orbits evolve. Larger protoplanets, represented by $n$-bodies, are tracked in detail \citep[e.g.,][]{kok1995, chambers2004,  raymond2004}, while smaller bodies that are too numerous to include as individual particles are accounted for statistically \citep[e.g.,][]{spaute1991, kl1998, weiden1997b} or with representative tracer particles \citep{lev2012, bk2020a}. The \orchestra\ software package \citep{kenyon2002, bk2006, kb2008, bk2011a, kb2016a}, a parallel hybrid coagulation~$+$~$n$-body code, enables cross-talk between the $n$-bodies, tracers and a statistical grid representing solids as small as submicron dust. For example, small particles orbitally damp larger ones by dynamical friction, facilitating accretion. The code implements this velocity evolution by shifting the eccentricity of the $n$-bodies, underscoring the importance of monitoring and modifying it accurately.

Between the various estimators of eccentricity described in \S\ref{sec:orbitchar} and the linearized theory of orbital dynamics around a central binary (Appendix~\ref{appx:linear}), we have the tools to simulate the velocity evolution of a swarm of $n$-body particles in a planet/satellite formation code. Our code uses osculating Keplerian orbital elements to track and adjust eccentricities when $\efree \gtrsim 0.02$, and linearized-theory for smaller eccentricities, where the epicyclic approximation is valid. In this way, we can damp (or excite) the $n$-body particles in a hybrid coagulation $+$ $n$-body simulation.

\begin{figure}[htb]
\centerline{\includegraphics[width=4.5in]{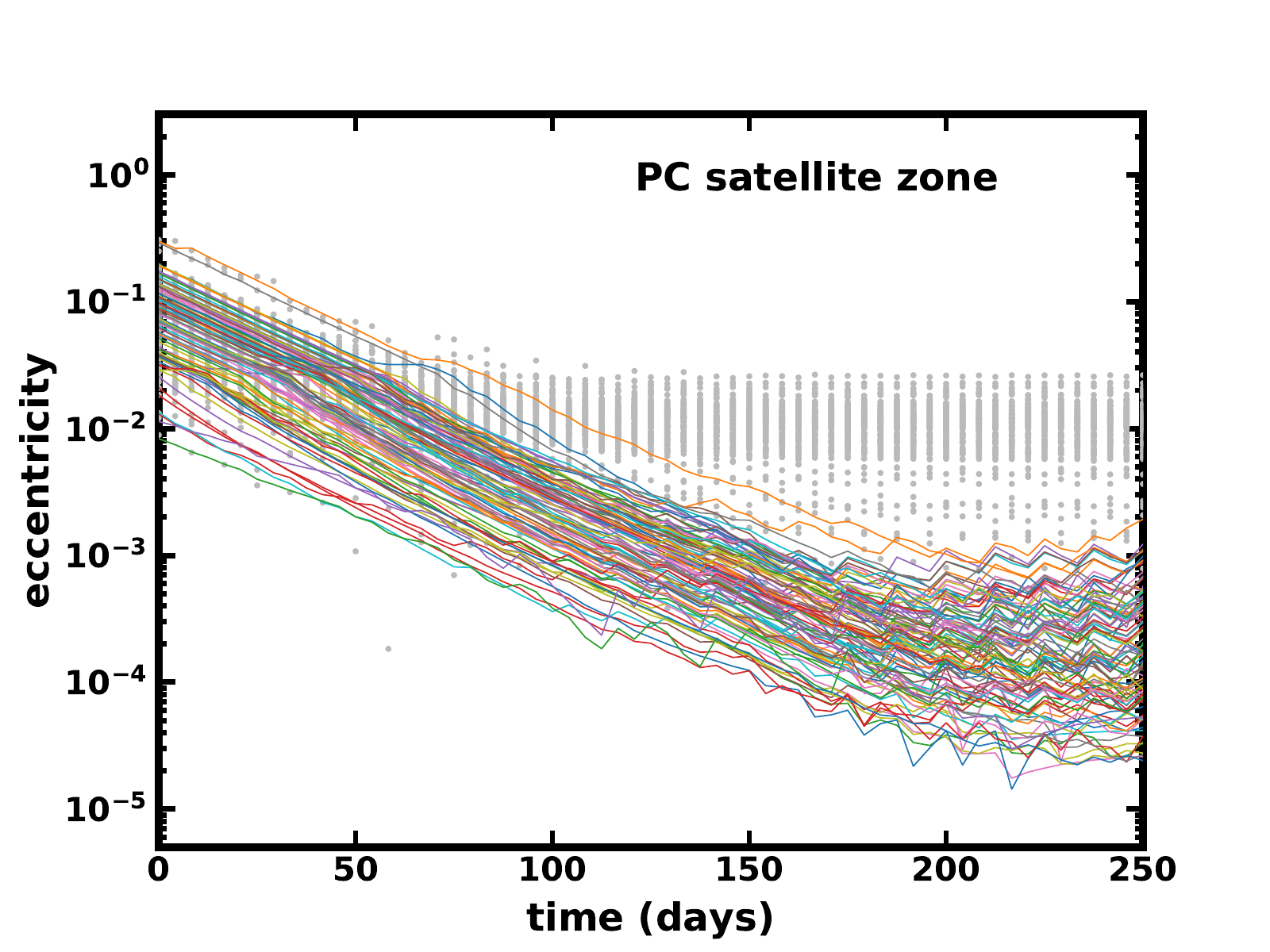}}
\caption{\label{fig:dampsim} 
    The evolution of eccentricity in a swarm of tracer particles in the satellite zone around the Pluto-Charon binary. Each line corresponds to a single tracer particle positioned randomly between the orbit of Styx and Hydra.  The gray dots are measurements of the Keplerian eccentricity. All tracers experience eccentricity damping at a rate of $de/dt = -C e$, where $C = 0.032$~yr$^{-1}$.  When eccentricities are high $\efree > 0.027$, we modify orbits using osculating Keplerian elements; when $\efree$ is low, we use linearized-theory. An artifact that results from this transition is evident in the plot.}
\end{figure}

Figure~\ref{fig:dampsim} provides an illustration. We place 100 tracer particles at random orbital distances from the Pluto-Charon barycenter within the satellite zone between Styx and Hydra.  Initially the tracers are on eccentric orbits, drawn from a Rayleigh distribution with an r.m.s.\ value of $e = 0.1$. As the particles evolve, their eccentricities are damped at a rate $de/dt \sim -C e$, where $C$ is a constant, set to be an identical, arbitrarily chosen value for all members of the swarm.  For $e\gtrsim 0.03$, eccentricities are monitored and modified in the Keplerian approximation ($\ekep$) with the Pluto-Charon binary treated as a point mass. At smaller eccentricities, linearized-theory is applied. The evolution stops when $\efree$  drops below the level at which $\eest$ can resolve a most-circular orbit (see Fig.~\ref{fig:eccvs}).  These eccentricities correspond to random speeds $\lesssim$0.5~m/s, well below the disruption threshold of weak ice ($\sim 4$ m/s for 5--10~m bodies).

For orbits around stellar binaries, the limitations of this approach for simulating velocity evolution are tied to the code's ability to resolve low-speed collisions. For example, Kepler-47 has a small binary eccentricity ($\ebin \approx 0.02$) and planets on orbits that are roughly 3.5 to 12 times the binary separation.  At the orbit of the innermost planet, Kepler-47b, the linearized-theory estimator $\eest$ can resolve $\efree \lesssim 2\times 10^{-4}$, which corresponds to a random speed below 20~m/s, smaller than the minimum collision speed required for the disruption of rocky bodies. At the orbital distance of the outermost planet, the code can resolve speeds as low as 20~cm/s.  However for Kepler-16b, with its more eccentric host, random speeds are only resolved to roughly 0.2~km/s, which is above the disruption threshold for rocky bodies (30 m/s for 200~m objects). If the code is to distinguish whether random motions lead to growth or fragmentation around Kepler-16, then other strategies for monitoring and adjusting eccentricities would be required.

\section{Conclusion}\label{sec:conclude}

Estimates of orbital elements of circumbinary satellites \citep[e.g.,][]{lee2006, showalter2015, woo2018, sutherland2019, woo2020} are essential to understanding satellite evolution \citep{smullen2017} and long-term stability \citep[e.g.,][]{kb2019b}. Here we explore how to characterize orbits using the linearized theory of \citet{lee2006} and \citet{leung2013}. Our goal is to find a fast and efficient way to monitor and modify eccentricities from orbital state vectors so that we can implement velocity evolution in $n$-body simulations of circumbinary planet formation.

We achieved this goal, but with some restrictions on the applicability of our method. The main quantitative tool, a measure of free eccentricity based on linearized theory (Eq.~(\ref{eq:eccest})), gives results that are most reliable when the orbital distance is large, beyond a few times the central binary separation, and when both the binary eccentricity and a satellite's free eccentricity are below about 0.1. Inclinations also must be similarly small.  These restrictions stem from the fact that the linearized theory is based on the epicyclic approximation of the orbits of the central binary and its satellites.

Despite these limitations, we identify systems where our method works well. In the satellite zone of the Pluto-Charon binary, we simulate the eccentricity damping of a dynamically hot swarm of tracer particles ($\efree \sim 0.1$) to a point where relative collision speeds drop below $\sim 10$~cm/s, roughly the escape speed from the surface of 100~m icy body. Thus, for the first time, we have simulated conditions in an $n$-body code where a swarm of circumbinary material is dynamically cool enough for coagulation to proceed. 

Our method is also applicable to the Kepler-47 planetary system. The low binary eccentricity ($e = 0.0234$) and comparatively large orbital distances of its planets suggest that our method can trace orbital damping down to collision speeds below about 10~m/s, well beneath the disruption threshold of rocky bodies.

Although our approach is reasonable for simulations with many particles, we do not advocate it for characterizing orbits from observational data.  More precise measures of orbital elements  include a geometric definition of eccentricity \citep[e.g.,][]{sutherland2019}, an FFT-based approach \citep{woo2018, woo2020}, and a fit to a Keplerian trajectory \citep{showalter2015}.  All require either multiple observations or numerical integration of an observed state vector.  Because eccentricity, at its essence, describes the shape of an orbital path, we recommend a geometry-based estimator, derived on the basis of radial excursions, so long as it accounts for the driving terms from the central binary's potential to isolate the free eccentricity (Eq.~(\ref{eq:egeo})).

Going forward, we plan to apply our method first to the Pluto-Charon system, working with the binary in its present, nearly-zero eccentricity configuration. The idea is to simulate the growth of the small satellites if they formed out of the dynamically hot debris from a giant collision between Charon and a Trans-Neptunian Object \citep{bk2020a}.  In this scenario, the impact occurred after the tidal expansion and circularization of the binary, thus avoiding sweeping resonances that might otherwise eject the small satellites.

We also hope to model planet formation around binary stars with this new method.  After all, Tatooine seems quite distant from its stellar hosts compared with the binary separation, which is just the right condition for our approach.

\acknowledgments We are grateful to M. Geller for comments on our manuscript. We also thank an anonymous referee for a report that led to improvements in the presentation. This work was support by NASA through Emerging Worlds program grant NNX17AE24G.

\appendix 
\section{Linearized theory details}\label{appx:linear}

Here we provide some mathematical details for deriving circumbinary orbit solutions. Our starting point, the gravitational potential at a satellite's position around a circular binary with primary and secondary masses $\Mp$ and $\Ms$, respectively, is
\begin{equation}\label{eq:Phi}
\Phi = -\frac{G \Mp}{[R^2+z^2+\Rp^2+2R \Rp \cos(\phi-\nbin t)]^{1/2}}
- \frac{G \Ms}{[R^2+z^2+\Rs^2-2R \Rs \cos(\phi-\nbin t)]^{1/2}},
\end{equation} 
where $(R,z,\phi)$ is the satellite's position in cylindrical coordinates with the origin at the center of mass, chosen so that the binary is in the $z=0$ plane. The positions of the primary and the secondary are specified by their radial distance from the origin, $\Rp$ and $\Rs$, along with their angular coordinates, $\phip = \nbin t +\pi$ and $\phis = \nbin t$, where $t$ is time and $\nbin$ is the mean motion of the binary. We choose $t = 0$ to denote when the satellite's guiding center and the secondary are aligned.

For simplicity we assume that satellites are in the plane of a circular binary, setting cylindrical coordinate $z=0$ and decompose Equation~(\ref{eq:Phi}) into harmonics of the
angle cosine of the satellite relative to the secondary \citep{lee2006}:
\begin{equation}\label{eq:Phiseries}
  \Phi \approx \sum_{k = 0}^{\infty}
  \Phi_{k}\, \cos(k\phi-k\nbin t))
  \ \ \ \ \ \ \ \ \ \ \ [k = 0,1,2,...],
\end{equation}
where the coefficients $\Phi_{k}$ depend on the binary masses, the binary separation $\abin$, and the orbital radius of the satellite's guiding center, $\rgc$. The form of these coefficients is
\begin{equation}\label{eq:Phik}
  \Phi_k =
\begin{dcases}
  -\frac{GM}{\rgc} \delta_{k0} -
  \frac{G\mu}{\rgc} \sum_{j=1} A^{(k)}_{j-k/2}
  \frac{{\Mfac}^{(+)}_{2j-1}}{M^{2j-1}}
    \frac{\abin^{2j}}{\rgc^{2j}} & (\text{even} \ k)
   \\
   -\frac{G\mu}{\rgc} \sum_{j=1} A^{(k)}_{j-(k-1)/2}
  \frac{{\Mfac}^{(-)}_{2j}}{M^{2j}}
    \frac{\abin^{2j+1}}{\rgc^{2j+1}} & (\text{odd} \ k)
\end{dcases}
\end{equation}
where $\delta_{k0}$ is the Kronicker delta function, $M=\Mp+\Ms$ is the total mass of the binary, $\mreduced = \Mp\Ms/M$ is the reduced mass, and the factors $A$ and $\cal{M}^{(\pm)}$ are
\begin{equation}
  A^{(k)}_{\ell} = \frac{2(2\ell)!(2k+2\ell)!}{2^{2k+4\ell}[\ell!(k+\ell)!]^2(1+\delta_{k0})}
  \ \ \ \text{and} \ \ \ \ {\Mfac}^{(\pm)}_{\ell} = \Mp^\ell \pm \Ms^\ell
  \ \ \ \ \ (\ell \geq 0);
\end{equation}
if the subscript $\ell < 0$, then both $A^{(k)}_\ell$ and ${\Mfac}^{(\pm)}_{\ell}$ are set to zero.\footnote{Our choice of lower limit of $j=1$ in the summations avoids unwanted terms for $k=0$ and $k=1$. That limit could otherwise be set higher, for example, to $j=k/2$ in the case of even $k$. For reference, the first few terms of $A^{(0)}_\ell$ are $[1,1/4,9/64,25/256,1225/16384]$ for $\ell=[0,1,2,3,4]$.}  These terms arise in a series expansion of Laplace coefficients from the spectral decomposition in Equation~(\ref{eq:Phiseries}) \citep[e.g.,][]{murray1999}.

The spectral decomposition of the binary potential reveals key aspects of a satellite's orbit.  The mean motion $\ngc$ and epicyclic frequency $\kappae$ are derivatives of the non-oscillatory part of the potential $\Phi_{0}$:
\begin{equation}\label{eq:ngc}
\ngc^2  \equiv \frac{1}{\rgc} \left.\frac{d\Phi_{00}}{dR}\right|_{\rgc}
  \frac{G M}{\rgc^3}\left\{1 + \frac{\mreduced}{M}\left[
    \frac{3}{4}\frac{\abin^2}{\rgc^2}  
     + \frac{45}{64}\frac{\Mfac^{(+)}_3}{M^3}\frac{\abin^4}{\rgc^4}
     + \frac{175}{256}\frac{\Mfac^{(+)}_5}{M^5}\frac{\abin^6}{\rgc^6}
     + ... \right]\right\},
\end{equation}
and
\begin{equation}\label{eq:kappae}
\kappae^2 \equiv \rgc \left.\frac{d\ngc^2}{dR}\right|_{\rgc}\!\!+4\ngc^2 = 
  \frac{G M}{\rgc^3}\left\{1 - \frac{\mreduced}{M}\left[
     \frac{3}{4}\frac{\abin^2}{\rgc^2}  
     + \frac{135}{64}\frac{\Mfac^{(+)}_3}{M^3}\frac{\abin^4}{\rgc^4}
     + \frac{875}{256}\frac{\Mfac^{(+)}_5}{M^5}\frac{\abin^6}{\rgc^6}
     + ...\right]\right\}.
\end{equation}
An the expansion of the potential in the $z$-direction similarly yields the vertical excursion frequency, $\kappai$.
\begin{equation}\label{eq:kappai}
\kappai^2 \equiv \left. \frac{1}{z}\frac{d\Phi}{dz}\right|_{z=0,\rgc}
= \frac{G M}{\rgc^3}\left\{1 + \frac{\mreduced}{M}\left[
     \frac{9}{4}\frac{\abin^2}{\rgc^2}
     + \frac{225}{64}\frac{\Mfac^{(+)}_3}{M^3}\frac{\abin^4}{\rgc^4}
     + \frac{1225}{256}\frac{\Mfac^{(+)}_5}{M^5}\frac{\abin^6}{\rgc^6}
     + ...\right]\right\}.
\end{equation}
From these quantities, we get apsidal and nodal precession rates, which characterize changes in a satellite's orbit over time scales much longer than a dynamical time \citep[see][]{mardling2013}. Equations (\ref{eq:ngc})--(\ref{eq:kappai}) hold for eccentric binaries as well \citep{leung2013}.

The frequencies described above all appear in the equations of motion in the main text (Equations~(\ref{eq:R})--(\ref{eq:z})); their harmonics give the driving frequencies of forced oscillations that stem from the relative motion of the binary and a satellite. The mode amplitudes are the coefficients $C_k$ and $D_k$, where
\begin{eqnarray}\label{eq:C}
  C_k & = & \left[\left.\frac{1}{\rgc}\frac{d\Phi_{k}}{dR}\right|_{\rgc}
  -\frac{2\ngc\Phi_{k}}{\rgc^2\nsyn}\right]\frac{1}{(\kappae^2-k^2\nsyn^2)}
\\ \label{eq:D}
  D_k & = & 2 C_k + \frac{\Phi_{k}}{\rgc^2\ngc\nsyn}.
\end{eqnarray}
These equations complete the analytical model for nearly coplanar orbits about a circular binary. \citet{leung2013} extend the theory to the more general case of an eccentric binary. 

\bibliography{planets}{}

\begin{thebibliography}{}
\expandafter\ifx\csname natexlab\endcsname\relax\def\natexlab#1{#1}\fi

\bibitem[{{Armstrong} {et~al.}(2014){Armstrong}, {Osborn}, {Brown}, {Faedi},
  {G{\'o}mez Maqueo Chew}, {Martin}, {Pollacco}, \& {Udry}}]{armstrong2014}
{Armstrong}, D.~J., {Osborn}, H.~P., {Brown}, D.~J.~A., {et~al.} 2014, \mnras,
  444, 1873

\bibitem[{{Benz} \& {Asphaug}(1999)}]{benz1999}
{Benz}, W., \& {Asphaug}, E. 1999, Icarus, 142, 5

\bibitem[{{Birnstiel} {et~al.}(2016){Birnstiel}, {Fang}, \&
  {Johansen}}]{birnstiel2016}
{Birnstiel}, T., {Fang}, M., \& {Johansen}, A. 2016, \ssr, 205, 41

\bibitem[{{Borucki} {et~al.}(2011){Borucki}, {Koch}, {Basri}, {Batalha},
  {Brown}, {Bryson}, {Caldwell}, {Christensen-Dalsgaard}, {Cochran}, {DeVore},
  {Dunham}, {Gautier}, {Geary}, {Gilliland}, {Gould}, {Howell}, {Jenkins},
  {Latham}, {Lissauer}, {Marcy}, {Rowe}, {Sasselov}, {Boss}, {Charbonneau},
  {Ciardi}, {Doyle}, {Dupree}, {Ford}, {Fortney}, {Holman}, {Seager},
  {Steffen}, {Tarter}, {Welsh}, {Allen}, {Buchhave}, {Christiansen}, {Clarke},
  {Das}, {D{\'e}sert}, {Endl}, {Fabrycky}, {Fressin}, {Haas}, {Horch},
  {Howard}, {Isaacson}, {Kjeldsen}, {Kolodziejczak}, {Kulesa}, {Li}, {Lucas},
  {Machalek}, {McCarthy}, {MacQueen}, {Meibom}, {Miquel}, {Prsa}, {Quinn},
  {Quintana}, {Ragozzine}, {Sherry}, {Shporer}, {Tenenbaum}, {Torres},
  {Twicken}, {Van Cleve}, {Walkowicz}, {Witteborn}, \& {Still}}]{borucki2011}
{Borucki}, W.~J., {Koch}, D.~G., {Basri}, G., {et~al.} 2011, \apj, 736, 19

\bibitem[{{Bromley} \& {Kenyon}(2006)}]{bk2006}
{Bromley}, B.~C., \& {Kenyon}, S.~J. 2006, \aj, 131, 2737

\bibitem[{{Bromley} \& {Kenyon}(2011)}]{bk2011a}
---. 2011, \apj, 731, 101

\bibitem[{{Bromley} \& {Kenyon}(2015)}]{bk2015tatooine}
---. 2015, \apj, 806, 98

\bibitem[{{Bromley} \& {Kenyon}(2020)}]{bk2020a}
---. 2020, \aj, 160, 85

\bibitem[{{Brozovi{\'c}} {et~al.}(2015){Brozovi{\'c}}, {Showalter}, {Jacobson},
  \& {Buie}}]{brozovic2015}
{Brozovi{\'c}}, M., {Showalter}, M.~R., {Jacobson}, R.~A., \& {Buie}, M.~W.
  2015, \icarus, 246, 317

\bibitem[{{Buie} {et~al.}(2006){Buie}, {Grundy}, {Young}, {Young}, \&
  {Stern}}]{buie2006}
{Buie}, M.~W., {Grundy}, W.~M., {Young}, E.~F., {Young}, L.~A., \& {Stern},
  S.~A. 2006, \aj, 132, 290

\bibitem[{{Chambers}(2004)}]{chambers2004}
{Chambers}, J.~E. 2004, Earth and Planetary Science Letters, 223, 241

\bibitem[{{Chambers}(2014)}]{chambers2014}
---. 2014, \icarus, 233, 83

\bibitem[{{Chavez} {et~al.}(2015){Chavez}, {Georgakarakos}, {Prodan},
  {Reyes-Ruiz}, {Aceves}, {Betancourt}, \& {Perez-Tijerina}}]{chavez2015}
{Chavez}, C.~E., {Georgakarakos}, N., {Prodan}, S., {et~al.} 2015, \mnras, 446,
  1283

\bibitem[{{Christy} \& {Harrington}(1978)}]{christy1978}
{Christy}, J.~W., \& {Harrington}, R.~S. 1978, \aj, 83, 1005

\bibitem[{{Doolin} \& {Blundell}(2011)}]{doolin2011}
{Doolin}, S., \& {Blundell}, K.~M. 2011, \mnras, 418, 2656

\bibitem[{{Doyle} {et~al.}(2011){Doyle}, {Carter}, {Fabrycky}, {Slawson},
  {Howell}, {Winn}, {Orosz}, {Prsa}, {Welsh}, {Quinn}, {Latham}, {Torres},
  {Buchhave}, {Marcy}, {Fortney}, {Shporer}, {Ford}, {Lissauer}, {Ragozzine},
  {Rucker}, {Batalha}, {Jenkins}, {Borucki}, {Koch}, {Middour}, {Hall},
  {McCauliff}, {Fanelli}, {Quintana}, {Holman}, {Caldwell}, {Still},
  {Stefanik}, {Brown}, {Esquerdo}, {Tang}, {Furesz}, {Geary}, {Berlind},
  {Calkins}, {Short}, {Steffen}, {Sasselov}, {Dunham}, {Cochran}, {Boss},
  {Haas}, {Buzasi}, \& {Fischer}}]{doyle2011}
{Doyle}, L.~R., {Carter}, J.~A., {Fabrycky}, D.~C., {et~al.} 2011, Science,
  333, 1602

\bibitem[{{Eriksson} {et~al.}(2020){Eriksson}, {Johansen}, \&
  {Liu}}]{eriksson2020}
{Eriksson}, L. E.~J., {Johansen}, A., \& {Liu}, B. 2020, \aap, 635, A110

\bibitem[{{Fleming} {et~al.}(2018){Fleming}, {Barnes}, {Graham}, {Luger}, \&
  {Quinn}}]{fleming2018}
{Fleming}, D.~P., {Barnes}, R., {Graham}, D.~E., {Luger}, R., \& {Quinn}, T.~R.
  2018, \apj, 858, 86

\bibitem[{{Georgakarakos} \& {Eggl}(2015)}]{georgakarakos2015}
{Georgakarakos}, N., \& {Eggl}, S. 2015, \apj, 802, 94

\bibitem[{{Goldreich} {et~al.}(2004){Goldreich}, {Lithwick}, \&
  {Sari}}]{gold2004}
{Goldreich}, P., {Lithwick}, Y., \& {Sari}, R. 2004, \araa, 42, 549

\bibitem[{{Holman} \& {Wiegert}(1999)}]{holman1999}
{Holman}, M.~J., \& {Wiegert}, P.~A. 1999, \aj, 117, 621

\bibitem[{{Huang} {et~al.}(2018){Huang}, {Shporer}, {Dragomir}, {Fausnaugh},
  {Levine}, {Morgan}, {Nguyen}, {Ricker}, {Wall}, {Woods}, \&
  {Vanderspek}}]{huang2018}
{Huang}, C.~X., {Shporer}, A., {Dragomir}, D., {et~al.} 2018, arXiv e-prints,
  arXiv:1807.11129

\bibitem[{{Ida} \& {Lin}(2004)}]{ida2004}
{Ida}, S., \& {Lin}, D.~N.~C. 2004, \apj, 604, 388

\bibitem[{{Johansen} {et~al.}(2014){Johansen}, {Blum}, {Tanaka}, {Ormel},
  {Bizzarro}, \& {Rickman}}]{johansen2014}
{Johansen}, A., {Blum}, J., {Tanaka}, H., {et~al.} 2014, Protostars and Planets
  VI, 547

\bibitem[{{Johansen} {et~al.}(2015){Johansen}, {Mac Low}, {Lacerda}, \&
  {Bizzarro}}]{johansen2015}
{Johansen}, A., {Mac Low}, M.-M., {Lacerda}, P., \& {Bizzarro}, M. 2015,
  Science Advances, 1, 1500109

\bibitem[{{Kennedy}(2015)}]{kennedy2015}
{Kennedy}, G.~M. 2015, \mnras, 447, L75

\bibitem[{{Kennedy} {et~al.}(2012){Kennedy}, {Wyatt}, {Sibthorpe}, {Phillips},
  {Matthews}, \& {Greaves}}]{kennedy2012}
{Kennedy}, G.~M., {Wyatt}, M.~C., {Sibthorpe}, B., {et~al.} 2012, \mnras, 426,
  2115

\bibitem[{{Kenyon}(2002)}]{kenyon2002}
{Kenyon}, S.~J. 2002, \pasp, 114, 265

\bibitem[{{Kenyon} \& {Bromley}(2008)}]{kb2008}
{Kenyon}, S.~J., \& {Bromley}, B.~C. 2008, \apjs, 179, 451

\bibitem[{{Kenyon} \& {Bromley}(2009)}]{kb2009}
---. 2009, \apjl, 690, L140

\bibitem[{{Kenyon} \& {Bromley}(2016)}]{kb2016a}
---. 2016, \apj, 817, 51

\bibitem[{{Kenyon} \& {Bromley}(2019)}]{kb2019b}
---. 2019, \aj, 158, 69

\bibitem[{{Kenyon} \& {Luu}(1998)}]{kl1998}
{Kenyon}, S.~J., \& {Luu}, J.~X. 1998, \aj, 115, 2136

\bibitem[{{Kley} \& {Haghighipour}(2015)}]{kley2015}
{Kley}, W., \& {Haghighipour}, N. 2015, \aap, 581, A20

\bibitem[{{Kokubo} \& {Ida}(1995)}]{kok1995}
{Kokubo}, E., \& {Ida}, S. 1995, Icarus, 114, 247

\bibitem[{{Kostov} {et~al.}(2013){Kostov}, {McCullough}, {Hinse}, {Tsvetanov},
  {H{\'e}brard}, {D{\'{\i}}az}, {Deleuil}, \& {Valenti}}]{kostov2013}
{Kostov}, V.~B., {McCullough}, P.~R., {Hinse}, T.~C., {et~al.} 2013, \apj, 770,
  52

\bibitem[{{Kostov} {et~al.}(2014){Kostov}, {McCullough}, {Carter}, {Deleuil},
  {D{\'{\i}}az}, {Fabrycky}, {H{\'e}brard}, {Hinse}, {Mazeh}, {Orosz},
  {Tsvetanov}, \& {Welsh}}]{kostov2014}
{Kostov}, V.~B., {McCullough}, P.~R., {Carter}, J.~A., {et~al.} 2014, \apj,
  784, 14

\bibitem[{{Kostov} {et~al.}(2016){Kostov}, {Orosz}, {Welsh}, {Doyle},
  {Fabrycky}, {Haghighipour}, {Quarles}, {Short}, {Cochran}, {Endl}, {Ford},
  {Gregorio}, {Hinse}, {Isaacson}, {Jenkins}, {Jensen}, {Kane}, {Kull},
  {Latham}, {Lissauer}, {Marcy}, {Mazeh}, {M{\"u}ller}, {Pepper}, {Quinn},
  {Ragozzine}, {Shporer}, {Steffen}, {Torres}, {Windmiller}, \&
  {Borucki}}]{kostov2016}
{Kostov}, V.~B., {Orosz}, J.~A., {Welsh}, W.~F., {et~al.} 2016, \apj, 827, 86

\bibitem[{{Kostov} {et~al.}(2020){Kostov}, {Orosz}, {Feinstein}, {Welsh},
  {Cukier}, {Haghighipour}, {Quarles}, {Martin}, {Montet}, {Torres}, {Triaud},
  {Barclay}, {Boyd}, {Briceno}, {Cameron}, {Correia}, {Gilbert}, {Gill},
  {Gillon}, {Haqq-Misra}, {Hellier}, {Dressing}, {Fabrycky}, {Furesz},
  {Jenkins}, {Kane}, {Kopparapu}, {Hod{\v{z}}i{\'c}}, {Latham}, {Law},
  {Levine}, {Li}, {Lintott}, {Lissauer}, {Mann}, {Mazeh}, {Mardling}, {Maxted},
  {Eisner}, {Pepe}, {Pepper}, {Pollacco}, {Quinn}, {Quintana}, {Rowe},
  {Ricker}, {Rose}, {Seager}, {Santerne}, {S{\'e}gransan}, {Short}, {Smith},
  {Standing}, {Tokovinin}, {Trifonov}, {Turner}, {Twicken}, {Udry},
  {Vanderspek}, {Winn}, {Wolf}, {Ziegler}, {Ansorge}, {Barnet}, {Bergeron},
  {Huten}, {Pappa}, \& {van der Straeten}}]{kostov2020}
{Kostov}, V.~B., {Orosz}, J.~A., {Feinstein}, A.~D., {et~al.} 2020, \aj, 159,
  253

\bibitem[{{Lambrechts} \& {Johansen}(2012)}]{lamb2012}
{Lambrechts}, M., \& {Johansen}, A. 2012, \aap, 544, A32

\bibitem[{{Lee} \& {Peale}(2006)}]{lee2006}
{Lee}, M.~H., \& {Peale}, S.~J. 2006, \icarus, 184, 573

\bibitem[{{Leinhardt} \& {Stewart}(2009)}]{lein2009}
{Leinhardt}, Z.~M., \& {Stewart}, S.~T. 2009, \icarus, 199, 542

\bibitem[{{Leung} \& {Lee}(2013)}]{leung2013}
{Leung}, G.~C.~K., \& {Lee}, M.~H. 2013, \apj, 763, 107

\bibitem[{{Levison} {et~al.}(2012){Levison}, {Duncan}, \& {Thommes}}]{lev2012}
{Levison}, H.~F., {Duncan}, M.~J., \& {Thommes}, E. 2012, \aj, 144, 119

\bibitem[{{Levison} {et~al.}(2015){Levison}, {Kretke}, \&
  {Duncan}}]{levison2015}
{Levison}, H.~F., {Kretke}, K.~A., \& {Duncan}, M.~J. 2015, \nat, 524, 322

\bibitem[{{Lines} {et~al.}(2014){Lines}, {Leinhardt}, {Paardekooper},
  {Baruteau}, \& {Thebault}}]{lines2014}
{Lines}, S., {Leinhardt}, Z.~M., {Paardekooper}, S., {Baruteau}, C., \&
  {Thebault}, P. 2014, \apjl, 782, L11

\bibitem[{{Lissauer} \& {Stewart}(1993)}]{liss1993b}
{Lissauer}, J.~J., \& {Stewart}, G.~R. 1993, in Protostars and Planets III, ed.
  E.~H. {Levy} \& J.~I. {Lunine}, 1061--1088

\bibitem[{{Lynden-Bell}(1963)}]{lyndenbell1963}
{Lynden-Bell}, D. 1963, The Observatory, 83, 23

\bibitem[{{Mardling}(2013)}]{mardling2013}
{Mardling}, R.~A. 2013, \mnras, 435, 2187

\bibitem[{{McKinnon} {et~al.}(2017){McKinnon}, {Stern}, {Weaver}, {Nimmo},
  {Bierson}, {Grundy}, {Cook}, {Cruikshank}, {Parker}, {Moore}, {Spencer},
  {Young}, {Olkin}, {Ennico Smith}, {New Horizons Geology}, {Imaging}, \&
  {Composition Theme Teams}}]{mckinnon2017}
{McKinnon}, W.~B., {Stern}, S.~A., {Weaver}, H.~A., {et~al.} 2017, \icarus,
  287, 2

\bibitem[{{Morbidelli}(2020)}]{morbidelli2020}
{Morbidelli}, A. 2020, \aap, 638, A1

\bibitem[{{Moriwaki} \& {Nakagawa}(2004)}]{mori2004}
{Moriwaki}, K., \& {Nakagawa}, Y. 2004, \apj, 609, 1065

\bibitem[{{Murray} \& {Dermott}(1999)}]{murray1999}
{Murray}, C.~D., \& {Dermott}, S.~F. 1999, {Solar system dynamics} (Princeton:
  Princeton University Press)

\bibitem[{{Nimmo} {et~al.}(2017){Nimmo}, {Umurhan}, {Lisse}, {Bierson},
  {Lauer}, {Buie}, {Throop}, {Kammer}, {Roberts}, {McKinnon}, {Zangari},
  {Moore}, {Stern}, {Young}, {Weaver}, {Olkin}, \& {Ennico}}]{nimmo2017}
{Nimmo}, F., {Umurhan}, O., {Lisse}, C.~M., {et~al.} 2017, \icarus, 287, 12

\bibitem[{{Ormel} \& {Klahr}(2010)}]{ormel2010a}
{Ormel}, C.~W., \& {Klahr}, H.~H. 2010, \aap, 520, A43

\bibitem[{{Orosz} {et~al.}(2012{\natexlab{a}}){Orosz}, {Welsh}, {Carter},
  {Fabrycky}, {Cochran}, {Endl}, {Ford}, {Haghighipour}, {MacQueen}, {Mazeh},
  {Sanchis-Ojeda}, {Short}, {Torres}, {Agol}, {Buchhave}, {Doyle}, {Isaacson},
  {Lissauer}, {Marcy}, {Shporer}, {Windmiller}, {Barclay}, {Boss}, {Clarke},
  {Fortney}, {Geary}, {Holman}, {Huber}, {Jenkins}, {Kinemuchi}, {Kruse},
  {Ragozzine}, {Sasselov}, {Still}, {Tenenbaum}, {Uddin}, {Winn}, {Koch}, \&
  {Borucki}}]{orosz2012a}
{Orosz}, J.~A., {Welsh}, W.~F., {Carter}, J.~A., {et~al.} 2012{\natexlab{a}},
  Science, 337, 1511

\bibitem[{{Orosz} {et~al.}(2012{\natexlab{b}}){Orosz}, {Welsh}, {Carter},
  {Brugamyer}, {Buchhave}, {Cochran}, {Endl}, {Ford}, {MacQueen}, {Short},
  {Torres}, {Windmiller}, {Agol}, {Barclay}, {Caldwell}, {Clarke}, {Doyle},
  {Fabrycky}, {Geary}, {Haghighipour}, {Holman}, {Ibrahim}, {Jenkins},
  {Kinemuchi}, {Li}, {Lissauer}, {Pr{\v s}a}, {Ragozzine}, {Shporer}, {Still},
  \& {Wade}}]{orosz2012b}
---. 2012{\natexlab{b}}, \apj, 758, 87

\bibitem[{{Pierens} \& {Nelson}(2007)}]{pierens2007}
{Pierens}, A., \& {Nelson}, R.~P. 2007, \aap, 472, 993

\bibitem[{{Pierens} \& {Nelson}(2013)}]{pierens2013}
---. 2013, \aap, 556, A134

\bibitem[{{Popova} \& {Shevchenko}(2013)}]{popova2013}
{Popova}, E.~A., \& {Shevchenko}, I.~I. 2013, \apj, 769, 152

\bibitem[{{Quarles} {et~al.}(2018){Quarles}, {Satyal}, {Kostov}, {Kaib}, \&
  {Haghighipour}}]{quarles2018}
{Quarles}, B., {Satyal}, S., {Kostov}, V., {Kaib}, N., \& {Haghighipour}, N.
  2018, \apj, 856, 150

\bibitem[{{Quintana} \& {Lissauer}(2006)}]{quintana2006}
{Quintana}, E.~V., \& {Lissauer}, J.~J. 2006, \icarus, 185, 1

\bibitem[{{Rafikov}(2013)}]{raf2013}
{Rafikov}, R.~R. 2013, \apjl, 764, L16

\bibitem[{{Raymond} {et~al.}(2004){Raymond}, {Quinn}, \&
  {Lunine}}]{raymond2004}
{Raymond}, S.~N., {Quinn}, T., \& {Lunine}, J.~I. 2004, Icarus, 168, 1

\bibitem[{{Ricker}(2015)}]{ricker2015}
{Ricker}, G.~R. 2015, in AAS/Division for Extreme Solar Systems Abstracts,
  Vol.~47, AAS/Division for Extreme Solar Systems Abstracts, 503.01

\bibitem[{{Safronov}(1969)}]{saf1969}
{Safronov}, V.~S. 1969, {Evoliutsiia doplanetnogo oblaka. (Evolution of the
  Protoplanetary Cloud and Formation of the Earth and Planets, Nauka, Moscow
  [Translation 1972, NASA TT F-677]} (1969.)

\bibitem[{{Schlichting}(2014)}]{schlichting2014}
{Schlichting}, H.~E. 2014, \apjl, 795, L15

\bibitem[{{Scholl} {et~al.}(2007){Scholl}, {Marzari}, \&
  {Th{\'e}bault}}]{scholl2007}
{Scholl}, H., {Marzari}, F., \& {Th{\'e}bault}, P. 2007, \mnras, 380, 1119

\bibitem[{{Schwamb} {et~al.}(2013){Schwamb}, {Orosz}, {Carter}, {Welsh},
  {Fischer}, {Torres}, {Howard}, {Crepp}, {Keel}, {Lintott}, {Kaib}, {Terrell},
  {Gagliano}, {Jek}, {Parrish}, {Smith}, {Lynn}, {Simpson}, {Giguere}, \&
  {Schawinski}}]{schwamb2013}
{Schwamb}, M.~E., {Orosz}, J.~A., {Carter}, J.~A., {et~al.} 2013, \apj, 768,
  127

\bibitem[{{Showalter} \& {Hamilton}(2015)}]{showalter2015}
{Showalter}, M.~R., \& {Hamilton}, D.~P. 2015, \nat, 522, 45

\bibitem[{{Showalter} {et~al.}(2011){Showalter}, {Hamilton}, {Stern}, {Weaver},
  {Steffl}, \& {Young}}]{showalter2011}
{Showalter}, M.~R., {Hamilton}, D.~P., {Stern}, S.~A., {et~al.} 2011, \iaucirc,
  9221, 1

\bibitem[{{Showalter} {et~al.}(2012){Showalter}, {Weaver}, {Stern}, {Steffl},
  {Buie}, {Merline}, {Mutchler}, {Soummer}, \& {Throop}}]{showalter2012}
{Showalter}, M.~R., {Weaver}, H.~A., {Stern}, S.~A., {et~al.} 2012, \iaucirc,
  9253, 1

\bibitem[{{Sigurdsson} {et~al.}(2003){Sigurdsson}, {Richer}, {Hansen},
  {Stairs}, \& {Thorsett}}]{sigurdsson2003}
{Sigurdsson}, S., {Richer}, H.~B., {Hansen}, B.~M., {Stairs}, I.~H., \&
  {Thorsett}, S.~E. 2003, Science, 301, 193

\bibitem[{{Simon} {et~al.}(2017){Simon}, {Armitage}, {Youdin}, \&
  {Li}}]{simon2017}
{Simon}, J.~B., {Armitage}, P.~J., {Youdin}, A.~N., \& {Li}, R. 2017, \apjl,
  847, L12

\bibitem[{{Smullen} \& {Kratter}(2017)}]{smullen2017}
{Smullen}, R.~A., \& {Kratter}, K.~M. 2017, \mnras, 466, 4480

\bibitem[{{Spaute} {et~al.}(1991){Spaute}, {Weidenschilling}, {Davis}, \&
  {Marzari}}]{spaute1991}
{Spaute}, D., {Weidenschilling}, S.~J., {Davis}, D.~R., \& {Marzari}, F. 1991,
  Icarus, 92, 147

\bibitem[{{Stern} {et~al.}(2015){Stern}, {Bagenal}, {Ennico}, {Gladstone},
  {Grundy}, {McKinnon}, {Moore}, {Olkin}, {Spencer}, {Weaver}, {Young},
  {Andert}, {Andrews}, {Banks}, {Bauer}, {Bauman}, {Barnouin}, {Bedini},
  {Beisser}, {Beyer}, {Bhaskaran}, {Binzel}, {Birath}, {Bird}, {Bogan},
  {Bowman}, {Bray}, {Brozovic}, {Bryan}, {Buckley}, {Buie}, {Buratti},
  {Bushman}, {Calloway}, {Carcich}, {Cheng}, {Conard}, {Conrad}, {Cook},
  {Cruikshank}, {Custodio}, {Dalle Ore}, {Deboy}, {Dischner}, {Dumont},
  {Earle}, {Elliott}, {Ercol}, {Ernst}, {Finley}, {Flanigan}, {Fountain},
  {Freeze}, {Greathouse}, {Green}, {Guo}, {Hahn}, {Hamilton}, {Hamilton},
  {Hanley}, {Harch}, {Hart}, {Hersman}, {Hill}, {Hill}, {Hinson}, {Holdridge},
  {Horanyi}, {Howard}, {Howett}, {Jackman}, {Jacobson}, {Jennings}, {Kammer},
  {Kang}, {Kaufmann}, {Kollmann}, {Krimigis}, {Kusnierkiewicz}, {Lauer}, {Lee},
  {Lindstrom}, {Linscott}, {Lisse}, {Lunsford}, {Mallder}, {Martin}, {McComas},
  {McNutt}, {Mehoke}, {Mehoke}, {Melin}, {Mutchler}, {Nelson}, {Nimmo},
  {Nunez}, {Ocampo}, {Owen}, {Paetzold}, {Page}, {Parker}, {Parker},
  {Pelletier}, {Peterson}, {Pinkine}, {Piquette}, {Porter}, {Protopapa},
  {Redfern}, {Reitsema}, {Reuter}, {Roberts}, {Robbins}, {Rogers}, {Rose},
  {Runyon}, {Retherford}, {Ryschkewitsch}, {Schenk}, {Schindhelm}, {Sepan},
  {Showalter}, {Singer}, {Soluri}, {Stanbridge}, {Steffl}, {Strobel}, {Stryk},
  {Summers}, {Szalay}, {Tapley}, {Taylor}, {Taylor}, {Throop}, {Tsang},
  {Tyler}, {Umurhan}, {Verbiscer}, {Versteeg}, {Vincent}, {Webbert}, {Weidner},
  {Weigle}, {White}, {Whittenburg}, {Williams}, {Williams}, {Williams},
  {Woods}, {Zangari}, \& {Zirnstein}}]{stern2015}
{Stern}, S.~A., {Bagenal}, F., {Ennico}, K., {et~al.} 2015, Science, 350,
  aad1815

\bibitem[{{Sutherland} \& {Kratter}(2019)}]{sutherland2019}
{Sutherland}, A.~P., \& {Kratter}, K.~M. 2019, \mnras, 487, 3288

\bibitem[{{van der Marel} {et~al.}(2015){van der Marel}, {Pinilla}, {Tobin},
  {van Kempen}, {Andrews}, {Ricci}, \& {Birnstiel}}]{vandermarel2015}
{van der Marel}, N., {Pinilla}, P., {Tobin}, J., {et~al.} 2015, \apjl, 810, L7

\bibitem[{{Weaver} {et~al.}(2016){Weaver}, {Buie}, {Buratti}, {Grundy},
  {Lauer}, {Olkin}, {Parker}, {Porter}, {Showalter}, {Spencer}, {Stern},
  {Verbiscer}, {McKinnon}, {Moore}, {Robbins}, {Schenk}, {Singer}, {Barnouin},
  {Cheng}, {Ernst}, {Lisse}, {Jennings}, {Lunsford}, {Reuter}, {Hamilton},
  {Kaufmann}, {Ennico}, {Young}, {Beyer}, {Binzel}, {Bray}, {Chaikin}, {Cook},
  {Cruikshank}, {Dalle Ore}, {Earle}, {Gladstone}, {Howett}, {Linscott},
  {Nimmo}, {Parker}, {Philippe}, {Protopapa}, {Reitsema}, {Schmitt}, {Stryk},
  {Summers}, {Tsang}, {Throop}, {White}, \& {Zangari}}]{weaver2016}
{Weaver}, H.~A., {Buie}, M.~W., {Buratti}, B.~J., {et~al.} 2016, Science, 351,
  aae0030

\bibitem[{{Weidenschilling} {et~al.}(1997){Weidenschilling}, {Spaute}, {Davis},
  {Marzari}, \& {Ohtsuki}}]{weiden1997b}
{Weidenschilling}, S.~J., {Spaute}, D., {Davis}, D.~R., {Marzari}, F., \&
  {Ohtsuki}, K. 1997, Icarus, 128, 429

\bibitem[{{Welsh} {et~al.}(2012){Welsh}, {Orosz}, {Carter}, {Fabrycky}, {Ford},
  {Lissauer}, {Pr{\v s}a}, {Quinn}, {Ragozzine}, {Short}, {Torres}, {Winn},
  {Doyle}, {Barclay}, {Batalha}, {Bloemen}, {Brugamyer}, {Buchhave},
  {Caldwell}, {Caldwell}, {Christiansen}, {Ciardi}, {Cochran}, {Endl},
  {Fortney}, {Gautier}, {Gilliland}, {Haas}, {Hall}, {Holman}, {Howard},
  {Howell}, {Isaacson}, {Jenkins}, {Klaus}, {Latham}, {Li}, {Marcy}, {Mazeh},
  {Quintana}, {Robertson}, {Shporer}, {Steffen}, {Windmiller}, {Koch}, \&
  {Borucki}}]{welsh2012}
{Welsh}, W.~F., {Orosz}, J.~A., {Carter}, J.~A., {et~al.} 2012, \nat, 481, 475

\bibitem[{{Welsh} {et~al.}(2015){Welsh}, {Orosz}, {Short}, {Cochran}, {Endl},
  {Brugamyer}, {Haghighipour}, {Buchhave}, {Doyle}, {Fabrycky}, {Hinse},
  {Kane}, {Kostov}, {Mazeh}, {Mills}, {M{\"u}ller}, {Quarles}, {Quinn},
  {Ragozzine}, {Shporer}, {Steffen}, {Tal-Or}, {Torres}, {Windmiller}, \&
  {Borucki}}]{welsh2015}
{Welsh}, W.~F., {Orosz}, J.~A., {Short}, D.~R., {et~al.} 2015, \apj, 809, 26

\bibitem[{{Wetherill}(1980)}]{weth1980}
{Wetherill}, G.~W. 1980, \araa, 18, 77

\bibitem[{{Wetherill} \& {Stewart}(1993)}]{weth1993}
{Wetherill}, G.~W., \& {Stewart}, G.~R. 1993, Icarus, 106, 190

\bibitem[{{Woo} \& {Lee}(2018)}]{woo2018}
{Woo}, J. M.~Y., \& {Lee}, M.~H. 2018, \aj, 155, 175

\bibitem[{{Woo} \& {Lee}(2020)}]{woo2020}
---. 2020, \aj, 159, 277

\bibitem[{{Youdin} \& {Kenyon}(2013)}]{yk2013}
{Youdin}, A.~N., \& {Kenyon}, S.~J. 2013, in Planets, Stars and Stellar
  Systems.~Volume 3: Solar and Stellar Planetary Systems, ed. T.~D. {Oswalt},
  L.~M. {French}, \& P.~{Kalas}, 1

\bibitem[{{Youdin} {et~al.}(2012){Youdin}, {Kratter}, \& {Kenyon}}]{youdin2012}
{Youdin}, A.~N., {Kratter}, K.~M., \& {Kenyon}, S.~J. 2012, \apj, 755, 17

\bibitem[{{Youdin} \& {Shu}(2002)}]{youdin2002}
{Youdin}, A.~N., \& {Shu}, F.~H. 2002, \apj, 580, 494

\end{thebibliography}

\end{document}